\documentclass{nature}
\usepackage{aas_macros,color,amsmath,amssymb}
\usepackage{graphicx,units,wasysym,hyperref}
\usepackage{lineno}
\usepackage{caption}
\def\xmm {XMM--Newton}
\def\nustar {NuSTAR}

\def\cxo {Chandra}
\def\swift {Swift}

\def\src {\mbox{NGC\,5907~ULX-1}}
\def\flux {\mbox{erg cm$^{-2}$ s$^{-1}$}}
\def\lum {\mbox{erg s$^{-1}$}}
\title{Diffuse X-ray emission around an ultraluminous X-ray pulsar}
\author{Andrea Belfiore,\!$^{1,\ast}$ Paolo Esposito,\!$^{2,1}$ 
    Fabio Pintore,\!$^{1}$ Giovanni Novara,\!$^{2}$
    Ruben Salvaterra,\!$^{1}$ Andrea De Luca,\!$^{1,3}$
    Andrea Tiengo,\!$^{2,1,3}$ Patrizia Caraveo,\!$^{1}$
    Felix F{\"u}rst,\!$^{4}$ Gian Luca Israel,\!$^{5}$ 
    Danilo Magistrali,\!$^{6}$ Martino Marelli,\!$^{1}$
    Sandro Mereghetti,\!$^{1}$ Alessandro Papitto,\!$^{5}$
    Guillermo A. Rodr\'iguez Castillo,\!$^{5}$ Chiara Salvaggio,\!$^{7,8}$ 
    Luigi Stella,\!$^{5}$ Dominic J. Walton,\!$^{9}$
    Anna Wolter,\!$^{8}$ Luca Zampieri$^{10}$
}
\begin{document}
%\linenumbers % this is required by nature

\maketitle

\begin{affiliations}
  \item {INAF--Istituto di Astrofisica Spaziale e Fisica Cosmica di Milano, 
      via A. Corti 12, 20133 Milano, Italy}
  \item {Scuola Universitaria Superiore IUSS Pavia, 
      piazza della Vittoria 15, 27100 Pavia, Italy}
  \item {INFN--Istituto Nazionale di Fisica Nucleare, Sezione di Pavia, 
      via A. Bassi 6, 27100 Pavia, Italy}
  \item {ESAC--European Space Astronomy Centre, Science Operations Departement,
      28692 Villanueva de la Ca{\~n}ada, Madrid, Spain}
  \item {INAF--Osservatorio Astronomico di Roma,
      via Frascati 33, 00078 Monteporzio Catone, Italy}
  \item {Universidad Pontificia Comillas Madrid--ICAI,
      Calle Alberto Aguilera 25, 28015 Madrid, Spain}
  \item {Department of Physics G. Occhialini, University of Milano--Bicocca,
      piazza dell'Ateneo Nuovo 1, 20126, Milano, Italy}
  \item {INAF--Osservatorio Astronomico di Brera,
      via Brera 28, 20121 Milano, Italy}
  \item {Institute of Astronomy, University of Cambridge,
      Madingley Road, Cambridge CB3 0HA, UK}
  \item {INAF--Osservatorio Astronomico di Padova,
      vicolo dell'Osservatorio 5, 35122 Padova, Italy}

$^{\ast}$e-mail: andrea.belfiore@inaf.it
  \end{affiliations}
%%%%%%%%%%%%%%%%%%%%%%%%%%%%%%%%%%%%%%%%%%%%%%%%%%%%%%%%%%%%%%%%%%%%%%%%%%%%%%%
\begin{abstract}
Ultraluminous X-ray sources (ULXs) are extragalactic X-ray emitters located
off-center of their host galaxy and with a luminosity in excess of a few
${10^{39}\text{\,\lum}}$, if emitted isotropically.\!\cite{feng11,kaaret17} 
The discovery of periodic modulation revealed that in  some ULXs the accreting
compact object is a neutron
star,\!\cite{bachetti14,furst16,israel17,ipe17,carpano18} indicating
luminosities substantially above their Eddington limit.
The most extreme object in this respect is \src\ (ULX1), with a peak
luminosity that is 500 times its Eddington limit.
During a \cxo\ observation to probe a low state of ULX1, we detected diffuse
X-ray emission at the position of ULX1.
Its diameter is $2.7 \pm 1.0$ arcsec and contains 25 photons, none below
0.8 keV.
We interpret this extended structure as an expanding nebula powered by 
the wind of ULX1.
Its diameter of about ${200\text{\,pc}}$, characteristic energy of
${\sim 1.9\text{\,keV}}$, and luminosity of ${\sim 2\times10^{38}\text{\,\lum}}$\ 
imply a mechanical power of ${1.3\times10^{41}\text{\,\lum}}$\ and an age
${\sim 7 \times 10^{4}\text{\,yr}}$.
This interpretation suggests that a genuinely super-Eddington regime can be 
sustained for time scales much longer than the spin-up time of the neutron
star powering the system.
As the mechanical power from a single ULX nebula can rival the injection
rate of cosmic rays of an entire galaxy,\!\cite{drury12} ULX nebulae 
could be important cosmic ray accelerators.\!\cite{abeysekara18}
\end{abstract}

\begin{figure}
%\internallinenumbers % required by nature
\resizebox{\hsize}{!}{\includegraphics{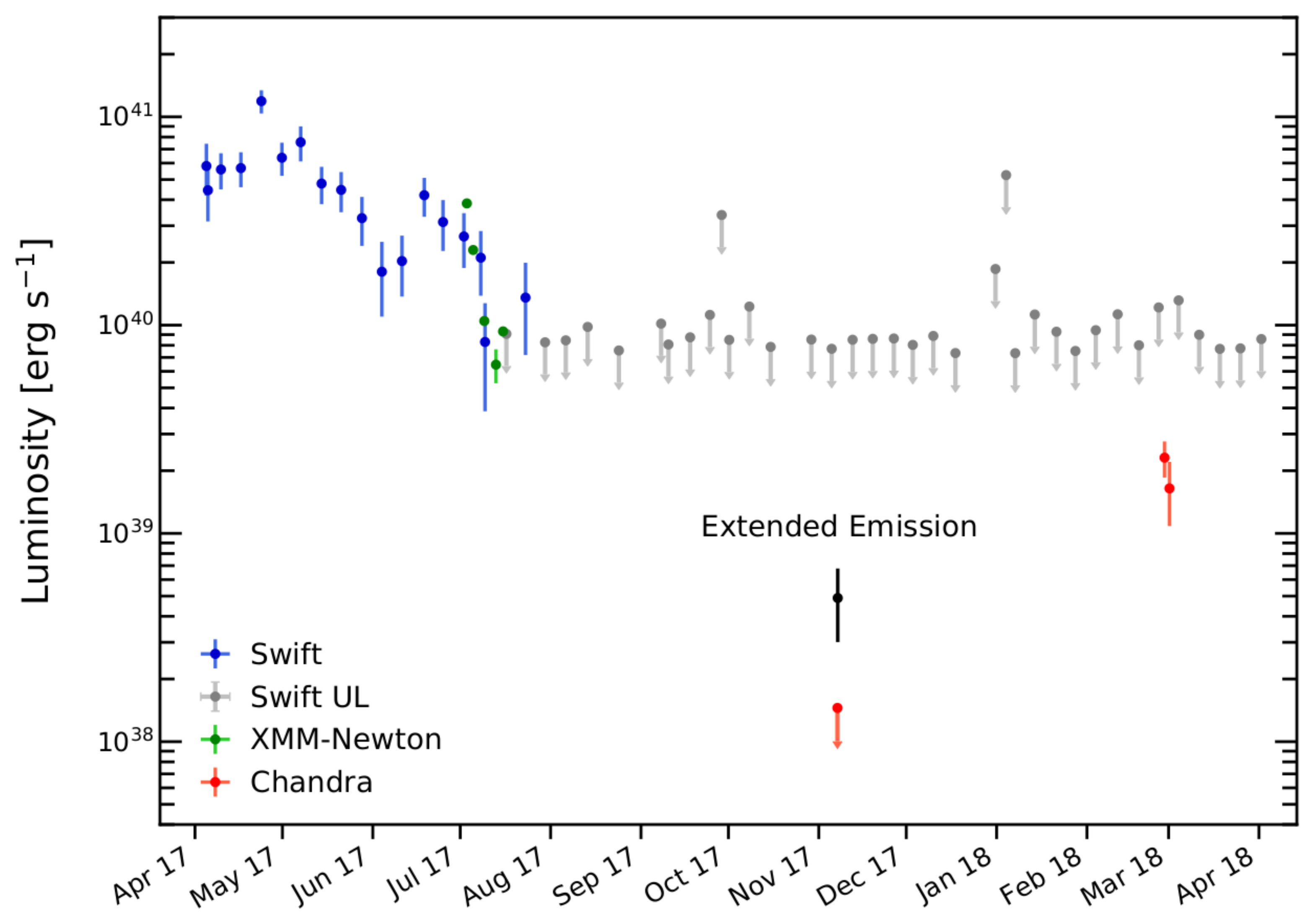}}
\caption{
\textbf{Multi-instrument soft X-ray light curve of \src\ since April 2017,
when the \swift\ monitoring resumed.}
The y-axis shows the luminosity in the  0.2--10\,keV energy range, assuming
a distance ${D\!=\!17.1\text{\,Mpc}}$.
In July 2017, when \xmm\ started observing, the source left its regularly 
modulated high state\cite{walton16}.
After ULX1 fell below the detection limit of \swift, \cxo\ observed
ULX1 in its low state in November 2017 and obtained an upper limit
(red downward arrow), assuming a power-law spectrum with index
${\Gamma\!=\!2}$.
The luminosity of the diffuse source (extended X-ray emission) associated
to ULX1 assumes a collisional plasma (apec) spectrum with 
$\text{k}_B T \!=\! 1.9_{-0.8}^{+2.3}\text{\,keV}$.
The grey arrows and blue points represent \swift\ upper limits and
detections respectively, assuming a broken power law spectrum, 
as modeled by \xmm\ in the high state of ULX1.\cite{israel17}
New \cxo\ observations (red points) taken in March 2018 found ULX1 
in an intermediate state.
All the error bars show uncertainties at the 90\% confidence level.}
\label{lightcurve}
\end{figure}

\begin{figure}
%\internallinenumbers % required by nature
\resizebox{\hsize}{!}{\includegraphics{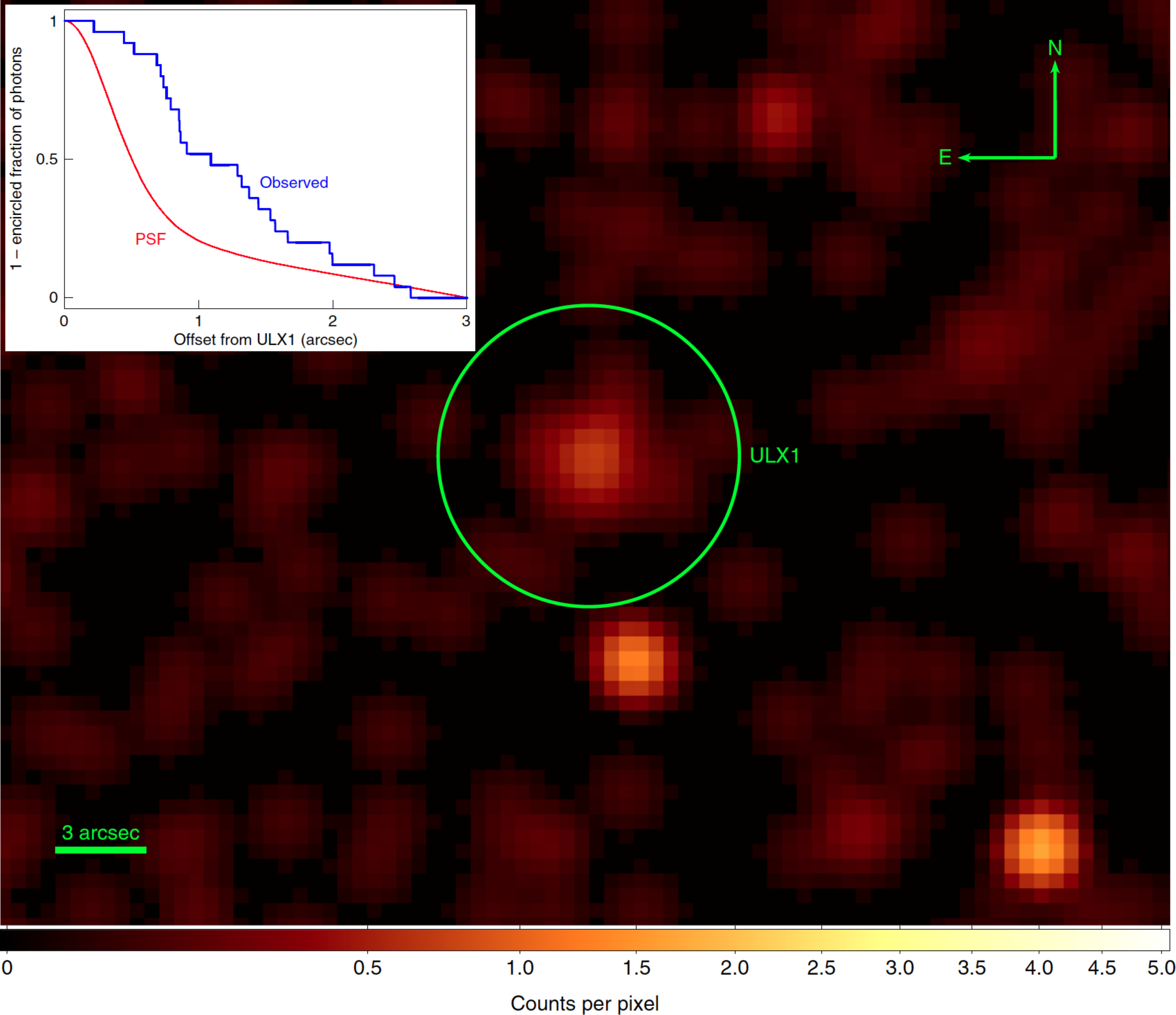}}
\caption{
\textbf{X-ray sky map between 0.3 and 7.0 keV of the region around the direction of
\src\ as observed by \cxo\ in November 2017.} 
Nearby sources are not significantly extended and their radial profile can be 
compared by eye with that of the diffuse emission. 
In the latter, no clear enhancement in brightness appears at the center of 
the source, indicating a discrepancy from the instrument PSF.
The green circle, with a radius of $5\text{\,arcsec}$, is centered on the
position of ULX1. 
The scale is in counts per pixel, whose side measures $0.5\text{\,arcsec}$,
after smoothing the image through a ${2-\text{D}}$ Gaussian kernel with
${\sigma_{\text{Gauss}}\!=\!1.5\text{\,pixel}}$.
The inset shows a comparison between the PSF (as obtained from simulations,
including background) and the observed radial distribution of the events (the
y-axis represents the fraction of events falling outside a certain radius).
We can reject the hypothesis that the source is point-like with a confidence 
level of 5$\sigma$ (${\text{p-value}\!=\!3\!\times\!10^{-7}}$).
See Fig.\,\ref{cxo_regions} for a broader and unsmoothed
version of this map.
}
\label{map}
\end{figure}

%%%%%%%%%%%%%%%%%%%%%%%%%%%%%%%%%%%%%%%%%%%%%%%%%%%%%%%%%%%%%%%%%%%%%%%%%%%%%%%
NGC\,5907 is a nearly edge-on (inclination of 87$^\circ$ 
[ref.\cite{xilouris99}])
spiral galaxy at a distance ${D\!=\!17.1\text{\,Mpc}}$ [ref.\cite{tully13}].
Its source \src\ (henceforth ULX1), with a peak X-ray luminosity
${\gtrsim 10^{41}\text{\,\lum}}$, is driven by an accreting neutron star
with a spin period of ${\sim \!1\text{\,s}}$ [ref.\cite{israel17}].
During an X-ray multi-instrument campaign to study its behavior, ULX1 dimmed
its flux by a factor ${>50}$ (Fig.\,\ref{lightcurve}) and, on
7 November 2017, \cxo\ observed the field of ULX1 with its Advanced CCD
Imaging Spectrometer for 52\,ks.
A source at the position of ULX1 was detected at a $7.7\sigma$ confidence 
level (Fig.\,\ref{map}).
Since the number of collected photons is rather low, we made extensive use of
simulations to assess the robustness of the findings and estimate the
parameters characterizing the source.
For our analysis, we considered the events between $0.3$ and $7.0\text{\,keV}$
falling within $3\text{\,arcsec}$ from the coordinates of ULX1, totalling 25
photons (3.6 of which are expected from background alone). 
The emission appears to be extended and indeed we verified that the photon
distribution in the detector is not consistent with the point-spread
function (PSF) of the telescope: we can reject a point-like nature of this
source at the $5\sigma$ confidence level.
We modelled the emission with a disk profile with uniform surface brightness
and estimated the disk radius to be ${R_\text{d}\!=\!1.35\pm0.50\text{\,arcsec}}$.
We found no indication of an excess of brightness at the center of the source. 
To set an upper limit on the flux of ULX1 (the point-like component of the
source), we selected the photons in the innermost part of the PSF, within
$0.5\text{\,arcsec}$, to compute a Poisson upper limit on the count rate.
Assuming a power-law spectrum with photon index $\Gamma=2$ and an absorbing 
column ${N_{\rm H}\!=\!5.3\!\times\!10^{21}\text{\,cm}^{-2}}$ (the value 
measured when the source was bright\cite{israel17}), we obtained an upper limit 
on the point source unabsorbed flux in the ${0.3-7.0\text{\,keV}}$ energy
range of $F_{\text{X}}\!<\!3.4\!\times\!10^{-15}$\,\flux\ at the 90\%
confidence level.
This corresponds to a limit on the isotropic X-ray luminosity
$L_{\text{X}}\!<\!1.2\!\times\!10^{38}$\,\lum, lower than the Eddington
luminosity ($L_{\text{Edd}} \sim 1.7\times\!10^{38}$\,\lum) for a
1.4-$M_{\odot}$ neutron star.

%%%%%%%%%%%%%%%%%%%%%%%%%%%%%%%%%%%%%%%%%%%%%%%%%%%%%%%%%%%%%%%%%%%%%%%%%%%%%%%
The diffuse emission at the position of ULX1 can be either an effect of the 
propagation of the high energy emission of the ULX through a thick dust layer 
in NGC\,5907 or the intrinsic emission of an extended structure physically 
associated to ULX1. 
X-ray dust-scattering could produce a halo of the size we observe only if 
ULX1 were located at least 6\,kpc behind the dominant scattering layer. 
However, in this case, given the high inclination and gas distribution of
NGC\,5907\cite{just06}, standard assumptions imply an X-ray absorption
along our line of sight much higher than the column density we derive
from the X-ray spectrum of ULX1.\!\cite{israel17}
(see the Supplementary  Information for a discussion and Fig.\,\ref{toy})
We therefore concentrate on the hypothesis of nebular emission surrounding
the ULX.
At the distance of NGC\,5907, the radius of the X-ray nebula is
${R\!=\!112\pm42\text{\,pc}}$. 
For a collisionally ionized diffuse gas and absorption 
${N_{\rm H}\!=\!(1.3^{+4.0}_{-1.2})\!\times\!10^{21}\text{\,cm}^{-2}}$
we estimate a temperature $T$ corresponding to
$k_B T\!=\!1.9_{-0.8}^{+2.3}$\,keV (where $k_B$ is the Boltzmann constant),
an unabsorbed flux 
${F_{\text{X}}\!=\!(6.1_{-2.5}^{+6.1})\!\times\!10^{-15}\text{\,\flux}}$\ 
and a luminosity
${L_{\text{X}}\!=\!(2.1_{-0.9}^{+2.1})\!\times\!10^{38}\text{\,\lum}}$.
% address to referee 2) start
A diffuse flux consistent with $L_{\text{Edd}}$ may suggest that the
Eddington-limited radiation from a neutron star is being Compton scattered
by the wind.
However, given that $\sigma_{\rm T}\times N_{\rm H} < 4\times 10^{-4}$
(where $\sigma_{\rm T}$ is the Thompson cross section), $>99\%$ of the
observed emission is unscattered.
% address to referee 2) end
If this were the remnant of a supernova, its energy would be
$E_{sn}>10^{53}\text{erg}$ [ref.\cite{blondin98}], corresponding to
a hypernova explosion. This is expected to produce a black hole\cite{fryer12}
but, in exotic scenarios, it could leave behind a neutron star.\cite{barkov11}

%%%%%%%%%%%%%%%%%%%%%%%%%%%%%%%%%%%%%%%%%%%%%%%%%%%%%%%%%%%%%%%%%%%%%%%%%%%%%%%
Nebular emission has been observed around several ULXs at optical and radio
wavelengths.\!\cite{pakull02,pakull03,pakull08,feng11,kaaret17,lang07,kaaret03}
These nebulae (often referred to as bubbles) are generally attributed to
shocks created by outflows from the binary system interacting with
the surrounding medium and show some common traits: diameter of
$\sim$200--400\,pc, expansion speed of $\sim$100--200\,km\,s$^{-1}$,
characteristic age (derived from the expansion velocity and the size
of the nebula) of $\sim$1\,Myr, and mechanical power
${\sim10^{39}\text{--}10^{40}\text{\,\lum}}$.\!
Here we explore whether a quasi-isotropic wind shocking the interstellar
medium (ISM)\cite{castor75,weaver77} can account for the extended X-ray
emission around ULX1. 
As the bubble expands with a radius $R\propto t^{\frac{3}{5}}$
[ref.\cite{castor75}],
the ISM accumulates just behind this external shock.
The wind, faster than the shock, also accumulates in another region closer
to the source.
Assuming that the X-ray emission comes from the outer region, pressure
equilibrium at the shock boundary provides an estimate of the shock velocity
${v_{\text{sh}}\!=\!\sqrt{\frac{5}{16}\frac{k_B T}{m_{\text{p}}}} \simeq 
1000\text{\,km\,s}^{-1}}$ (where $k_B T$ comes from the spectral fit 
and $m_{\text{p}}$ is the proton mass), which entails an age of the bubble
$\tau\!=\!\frac{3}{5}\frac{R}{v_{\text{sh}}}\!=\!
    \left( 6.7_{-2.8}^{+3.1} \right)\!\times\!10^{4}\text{\,yr}$.
This value is much larger than the spin-up timescale of the neutron
star that powers the system, $40\,\text{yr}$ [ref.\cite{israel17}].
We also derived an estimate of the ISM density (see Methods) as 
${n_{\text{ISM}} \simeq 0.08\text{\,cm}^{-3}}$ and of the mechanical power
carried by the wind, 
${L_{\text{w}}\!=\!\left( 1.3_{-1.0}^{+9.8} \right)\!\times\!10^{41}\text{\,\lum}}$,
not far from the value of ${5\!\times\!10^{40}\text{\,\lum}}$\ estimated 
for the bubble S26 in NGC\,7793.\cite{pakull10,dopita12}
If this mechanical power were sustained for ${\sim 7 \times 10^4\text{\,yr}}$,
then, assuming a typical accretion efficiency onto a neutron star of 17\%,
${\sim 0.9\,M_\odot}$ should have been accreted to provide enough
energy to sustain the nebula.
Because the wind carries mechanical power to the nebula, if a large mass
has been accreted onto the neutron star, and the mass lost by the system cannot
exceed a few 10\,$M_{\odot}$, then we obtain a speed of the wind 
$v_{\text{w}}\!\apprge\!0.1\,c$ where $c$ is the speed of light.
This value is consistent with the outflows observed from other
ULXs.\cite{pinto16,kosec18,walton16b}

NGC 5907 has been observed in ${\text{H}_\alpha}$ with the Kitt Peak
National Observatory ${0.9\text{ m}}$ telescope in May 1995.\cite{rand96}
The high level of contamination from star forming regions and the limited
angular resolution (${\sim 1\text{\,arcsec}}$) hamper a detection
of a counterpart in ${\text{H}_\alpha}$ to the nebula around ULX1.
\src\ has been observed in radio at $5\text{ GHz}$  with the Very Large
Array in May 2012,\cite{mezcua13} detecting no point source down to a flux
density of ${20 \text{ } \mu \text{Jy}}$.
The radio emission expected from the hot X-ray-emitting plasma is much
fainter than this limit (see the Supplementary Information for a discussion).
However, because efficient radiative cooling would have boosted the radio
emission to ${\sim 240 \text{ } \mu \text{Jy}}$[ref.\cite{dopita96,caplan86}],
this limit confirms our adiabatic approximation and justifies the large ratio
${L_{\mathrm{w}} / L_{\mathrm{X}}}$.

%%%%%%%%%%%%%%%%%%%%%%%%%%%%%%%%%%%%%%%%%%%%%%%%%%%%%%%%%%%%%%%%%%%%%%%%%%%%%%%
Observing an X-ray bubble around a ULX requires a number of favourable
circumstances: a bubble with the right surface brightness and size to be
detectable as extended; a ULX dim enough not to outshine the bubble;
a sensitive observation carried out with an instrument with good angular
resolution (see the Supplementary Information for a quantitative discussion
and Fig.\,\ref{detectability}).
This might explain why no similar structures are commonly observed
in association with ULXs (with the notable exception of 
NGC\,7793\,S26,\cite{pakull10} which, however, has no associated ULX).
Indeed, two follow-up observations with \cxo\ performed between
2018 February 27 and March 1 for a combined exposure of 50\,ks failed to
detect the bubble, as ULX1 raised its luminosity to
$4\!\times\!10^{39}$\,\lum. 

The recent discovery of TeV emission from the Galactic microquasar
SS\,433,\cite{abeysekara18} in many ways reminiscent of
ULXs,\cite{begelman06} suggests that strong shocks associated to ULXs 
contribute to the cosmic ray acceleration.
Indeed, the mechanical power of the bubble of ULX1 is comparable to the
cosmic-ray injection rate for a whole galaxy.\cite{drury12}
Since the duty cycle, the lifespan and the population of similar objects 
are currently poorly known, it is not possible at this stage to quantify 
their contribution. 

%%%%%%%%%%%%%%%%%%%%%%%%%%%%%%%%%%%%%%%%%%%%%%%%%%%%%%%%%%%%%%%%%%%%%%%%%%%%%%%

%%%%%%%%%%%%%%%%%%%%%%%%%%%%%%%%%%%%%%%%%%%%%%%%%%%%%%%%%%%%%%%%%%%%%%%%%%%%%%%

\clearpage

\noindent{\large\bf Methods}\\
%%%%%%%%%%%%%%%%%%%%%%%%%%%%%%%%%%%%%%%%%%%%%%%%%%%%%%%%%%%%%%%%%%%%%%%%%%%%%%%
\noindent\textbf{X-ray observations and long-term light curve}\\
We consider in this work X-ray observations taken with \cxo, \xmm, and the
Neil Gehrels Swift Observatory (see Table\,\ref{obslog}).
Particular care is dedicated to the \cxo\ observation 20830, taken on 2017
November 6, in which we detected diffuse emission.
We used the other data sets to build the long-term light curve of
Fig.\,\ref{lightcurve} and/or to improve the astrometry of observation 20830.

%%%%%%%%%%%%%%%%%%%%%%%%%%%%%%%%%%%%%%%%%%%%%%%%%%%%%%%%%%%%%%%%%%%%%%%%%%%%%%%
\begin{table}
%\internallinenumbers % required by nature
\centering
\begin{tabular}{llcccc}
\hline
\# & Mission & Obs.ID & Instrument & Start Date & Total Exposure\\
 & & & & (YYYY-MM-DD) & (ks) \\
\hline
1 & \cxo & 12987 & ACIS-S & 2012-02-11 & 16.0 \\
2 & \xmm & 0804090301 & EPIC & 2017-07-02 & 43.0 (28.0) \\
3 & \xmm & 0804090401 & EPIC & 2017-07-05 & 39.0 (31.3) \\
4 & \xmm & 0804090501 & EPIC & 2017-07-08 & 43.0 (34.8) \\
5 & \xmm & 0804090701 & EPIC & 2017-07-12 & 43.0 (32.5) \\
6 & \xmm & 0804090601 & EPIC & 2017-07-15 & 40.5 (24.8) \\
7 & \cxo & 20830 & ACIS-S & 2017-11-07 & 51.3 \\
8 & \cxo & 20994 & ACIS-S & 2018-02-27 & 32.6 \\
9 & \cxo & 20995 & ACIS-S & 2018-03-01 & 16.0 \\
\hline \\
\end{tabular}
\caption{
\textbf{Log of the \xmm\ and \cxo\ observations used in this work.}
The last column reports the duration of each observation and, for \xmm,
in parenthesis, also the net PN exposure, after filtering for background
flares. \label{obslog}}
\end{table}

\begin{figure}
%\internallinenumbers % required by nature
\centering
\resizebox{\hsize}{!}{\includegraphics{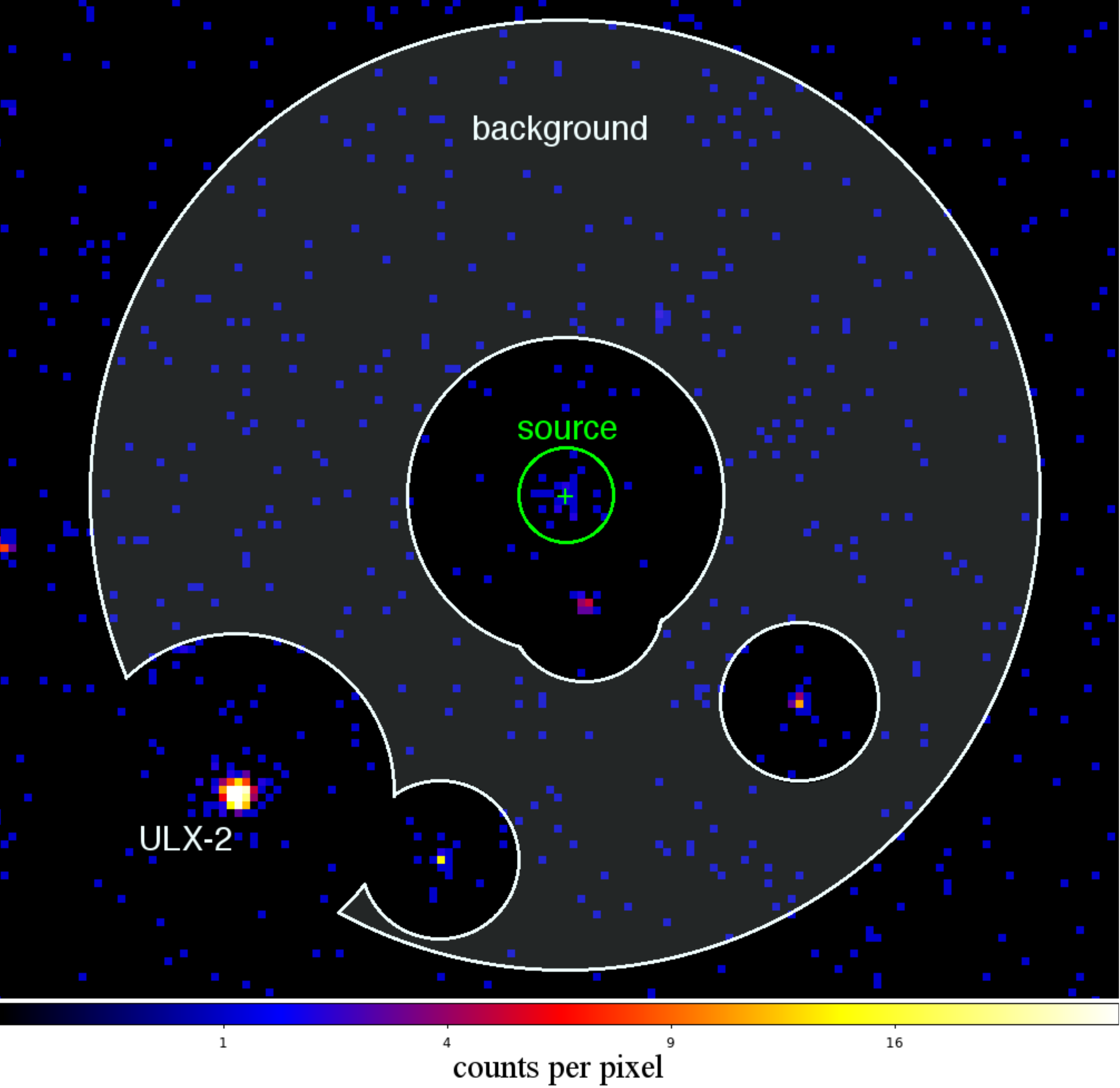}}
\caption{
\textbf{Counts map between 0.3 and 7.0 keV from the \cxo\ observation
of November 2017 (Obs.~Id.~20830), centered on the direction of ULX1.}
ULX1 is marked here by a green cross. 
The green circle represents the source extraction region and contains 25 photons. 
The white shaded area covers the region used to estimate the background level
and contains 277 photons.
It is shaped as an annulus centered on ULX1 from which circles around each
source have been removed.
We expect a background contamination of 3.6 photons in the source region.
The bright source in the bottom left corner is NGC 5907 ULX2.
}
\label{cxo_regions}
\end{figure}

%%%%%%%%%%%%%%%%%%%%%%%%%%%%%%%%%%%%%%%%%%%%%%%%%%%%%%%%%%%%%%%%%%%%%%%%%%%%%%%
\noindent\emph{Chandra}\\
All \cxo\ observations we used were carried out with the Advanced
CCD Imaging Spectrometer\cite{garmire03} (Spectroscopic array, ACIS-S) in
full-imaging mode.
We used the Chandra Interactive Analysis of Observation (CIAO,
v4.10)\cite{fruscione06} software package and CALDB (v4.7.8).
We selected photons in the energy range 0.3--7\,keV, and followed the standard
analysis threads for data reprocessing, source detection, and flux estimation.
We determined the energy boundaries based on the spectral distribution
of the ACIS background, which increases significantly outside this band.
We defined a circular source region that includes all photons within
${3\text{\,arcsec}}$ from the direction of ULX1.
This radius, which contains 98.5\% of the PSF, limits the contamination
from background and other sources, but is large enough to make it possible
to appreciate the main features of the X-ray source.
We defined an annular background region centered at the location of ULX1,
with radii $10\text{\,arcsec}$ and $30\text{\,arcsec}$.
We removed from this region circles with radii of $5\text{\,arcsec}$
around each other source, and a circle with radius of $10\text{\,arcsec}$
around NGC\,5907\,ULX-2, being particularly bright at the time. 
For observation 20830, these criteria leave 25 photons in the source region
and 277 photons in the background region, corresponding to 3.6 background
photons expected in the source region (see Fig.\,\ref{cxo_regions}).

%%%%%%%%%%%%%%%%%%%%%%%%%%%%%%%%%%%%%%%%%%%%%%%%%%%%%%%%%%%%%%%%%%%%%%%%%%%%%%%

%%%%%%%%%%%%%%%%%%%%%%%%%%%%%%%%%%%%%%%%%%%%%%%%%%%%%%%%%%%%%%%%%%%%%%%%%%%%%%%
\noindent\emph{XMM--Newton}\\
\noindent In the \xmm\ observations, the positive\textendash negative
junction\cite{struder01} (pn) and the two metal oxide 
semi-conductor\cite{turner01} (MOS) CCD cameras of the EPIC instrument were 
all operated in Full Frame mode. 
We used the \xmm\ Science Analysis Software\cite{gabriel04} (SAS) v14.5 
for data reduction.
After removing intervals of high background, we selected the events setting 
FLAG==0 and PATTERN$<=$4 and PATTERN$<=$12 for pn and MOS, respectively.
We extracted the source spectra and event lists from a circular region with
radius 30\,arcsec around the best-fit \cxo\ source position, 
${\text{RA}\!=\!228.994289^\circ}$  and 
${\text{Dec}\!=\!56.302851^\circ}$ (J2000), in the energy range 0.3--10\,keV.
We estimated the background from a circular region with radius 65\,arcsec,
close to ULX1 but free of sources, for each observation.
We excluded from our analysis the \xmm\ observation 0795712601 as source
contamination, mainly from NGC\,5907\,ULX-2 but also from other sources
in NGC\,5907, undermines a clear characterization of ULX1.\cite{pintore18}

%%%%%%%%%%%%%%%%%%%%%%%%%%%%%%%%%%%%%%%%%%%%%%%%%%%%%%%%%%%%%%%%%%%%%%%%%%%%%%%
We simultaneously analysed all \xmm\ spectra in the energy range 
0.3--10\,keV and we fitted them with an absorbed broken power law
model (bknpow in XSPEC\cite{arnaud96}). The Tuebingen-Boulder ISM 
absorption model (tbabs) was adopted and the abundances were set to
those of ref.\cite{wilms00}.
We fixed the values of the break energy and the high-energy spectral index,
to those obtained in ref.\,\cite{israel17},
(${E_\text{b}\!=\!6.7\text{\,keV}}$ and ${\Gamma_{\!2}\!=\!2.9}$), as they
were better constrained including \nustar\ data and consistent with all
the \xmm\ observations of ULX1 in a high state.
We also fixed the column density to the best-fit value in 
ref.\,\cite{israel17}, ${N_{\rm H}\!=\!5.3\!\times\!10^{21}\text{\,cm}^{-2}}$.
We left free to vary the low-energy photon index and the normalization.
We obtained an acceptable fit (${\chi^2_{\nu}\!=\!1.12}$ for 439 degrees of
freedom), with low-energy photon indices of
${\Gamma_{\!1}^{(2)}\!\!=\!1.72\pm0.03}$,
${\Gamma_{\!1}^{(3)}\!\!=\!1.88\pm0.04}$, 
${\Gamma_{\!1}^{(4)}\!\!=\!1.90\pm0.06}$,
${\Gamma_{\!1}^{(5)}\!\!=\!2.82\pm0.18}$,
and ${\Gamma_{\!1}^{(6)}\!\!=\!2.08\pm0.07}$
for the five observations, in chronological order (the superscripts
refer to the observation codes in Table\,\ref{obslog}).
The resulting 0.3--10\,keV unabsorbed fluxes are 
${F_\text{X}^{(2)}\!\!=\!(7.09\pm0.15)\!\times\! 10^{-13}\text{\,\flux}}$,
${F_\text{X}^{(3)}\!\!=\!(4.22\pm0.10)\!\times\! 10^{-13}\text{\,\flux}}$,\hfill\break
${F_\text{X}^{(4)}\!\!=\!(1.92\pm0.07)\!\times\! 10^{-13}\text{\,\flux}}$,
${F_\text{X}^{(5)}\!\!=\!(9.2\pm1.2)\!\times\! 10^{-14}\text{\,\flux}}$, and \hfill\break
${F_\text{X}^{(6)}\!\!=\!(1.65\pm0.07)\!\times\! 10^{-13}\text{\,\flux}}$
(see Fig.\,\ref{lightcurve}).
All these uncertainties are at the 90\% confidence level.
A careful spectral characterization, to be compared with other spectral
analyses in the literature, goes beyond the scope of this paper.

%%%%%%%%%%%%%%%%%%%%%%%%%%%%%%%%%%%%%%%%%%%%%%%%%%%%%%%%%%%%%%%%%%%%%%%%%%%%%%%
\noindent\emph{Swift}\\
\noindent The X-Ray Telescope\cite{burrows05} (XRT) on board \swift\ uses 
a CCD detector  sensitive to photons with energies between 0.2 and 10\,keV.
All observations analyzed in this work were performed in imaging photon
counting (PC) mode.
We used FTOOLS\cite{blackburn95} v6.15 for standard data processing. 
We extracted the source events within a radius of $20\text{\,arcsec}$ 
from the \cxo\ position of ULX1, and evaluated the background in a 
source-free circular region of radius $130\text{\,arcsec}$, avoiding 
the plane of NGC\,5907. 
The ancillary response files generated with xrtmkarf account for different
extraction regions, vignetting and point-spread function corrections.
We used the latest available spectral redistribution matrix (v014).
We converted the source rate to 0.2--10\,keV luminosity by assuming
a distance ${D\!=\!17.1\text{\,Mpc}}$ and an absorbed broken power-law spectral
model with indices 1.57 and 2.87, and break energy ${6.42\,\text{keV}}$.
These are the best-fit parameters obtained in the high state of
ULX1,\cite{israel17} consistent with, but more constrained, than the values
obtained in the spectral analysis described in the previous section.

As the region around ULX1 is rich of X-ray sources from NGC\,5907, which
\cxo\ can resolve but \swift\ cannot, XRT observations are likely affected
by source contamination.
A dedicated analysis that addresses this issue is not straightforward 
and goes beyond the scope of this paper.

%%%%%%%%%%%%%%%%%%%%%%%%%%%%%%%%%%%%%%%%%%%%%%%%%%%%%%%%%%%%%%%%%%%%%%%%%%%%%%%
\noindent\textbf{Analysis of the \cxo\ observation of diffuse emission around
ULX1}\\
\noindent\emph{Relative astrometry}\label{astrometry}\\
We considered \cxo\ observations 12987 and 20994, besides observation 20830.
We reprocessed the data, extracted images, and run the CIAO task wavdetect, 
following the indications in the \cxo\ analysis threads. 
We selected the sources within 1\,arcmin from the nominal position of ULX1, 
excluding ULX1 itself.
We found the translations that best map the coordinates of the sources 
in observations 12987 and 20994 to those in observation 20830, using the CIAO 
task wcs\_match. 
We applied these corrections (measuring 0.13\,arcsec and 0.30\,arcsec,
respectively) to the images and aspect solutions of observations
12987 and 20994 and launched again the task wavdetect. 
The localization of ULX1 in the two observations is now compatible within
$1\sigma$ in RA 
(${\left|\Delta \text{RA}\right|\!=\!0.06\text{\,arcsec}}$) and Dec
(${\left|\Delta \text{Dec}\right|\!=\!0.06\text{\,arcsec}}$), in the 
relative frame of observation 20830. 
Therefore, we take the barycenter of these two positions as the (J2000)
nominal position of ULX1: ${\text{RA}\!=\!228.994289(8)^\circ}$ and 
${\text{Dec}\!=\!56.302851(4)^\circ}$ (or 
${\text{RA}\!=\!15^\text{h}15^\text{m}58.\!\!^\text{s}63}$,
${\text{Dec}\!=\!+56^\circ18'10.\!\!{''}3}$). 
The best fit coordinates of the closest source to ULX1 in observation 20830 
are: ${\text{RA}\!=\!228.99414(11)^\circ}$ and
${\text{Dec}\!=\!56.30284(6)^\circ}$ (J2000). 
These two positions are compatible within $2\sigma$, being offset by 
${0.3\text{\,arcsec}}$, with a $1\sigma$ uncertainty of 
${0.21\text{\,arcsec}}$. 
Therefore, we identify the source in observation 20830 with ULX1 and adopt 
the nominal position of ULX1 in the analysis that follows.

%%%%%%%%%%%%%%%%%%%%%%%%%%%%%%%%%%%%%%%%%%%%%%%%%%%%%%%%%%%%%%%%%%%%%%%%%%%%%%%
\noindent\emph{Extension}\\
A first indication of source extension comes from the output of the CIAO tool
wavdetect. 
We tested for and measured the extension of our source, by studying its
radial brightness profile.
We used a Kolmogorov--Smirnov test, which quantifies the goodness of fit by 
measuring the maximum absolute difference between the cumulative distribution
of observed events and the model. 
To take into account the complex shape of the PSF and other effects due to 
the spacecraft dithering, we simulated with MARX\cite{davis12} (v5.3.3) a 
large number of events associated to a point source with characteristics 
similar to ULX1 in observation 20830, including its position on the detector.
In order to avoid photon pileup, which alters the PSF, we generated 50 
realizations of a source with 1000 events and combined them. 
We assumed an absorbed ($N_{\rm H}\!=\!5.3\!\times\!10^{21}$\,cm$^{-2}$,
the value measured for ULX1 in the high state\cite{israel17}), power-law 
($\Gamma\!=\!2$) spectrum and a direction consistent with those of ULX1.
We obtain consistent results assuming the best-fit thermal bremsstrahlung
spectral model described in the next section.
We generated a number of simulated background events, drawing from
a two-dimensional uniform distribution.

We extracted the observed cumulative distribution of events within a radius $R$
from the source center. 
We used the nominal direction of ULX1 obtained through astrometric analysis 
(see the previous section), but using the observed centroid does not alter our 
results. 
For the simulated point source, we included the expected uniform background
contribution. 
We applied the same selection criteria to the simulated data set and
considered only photons in the source region (with radius 3\,arcsec):
25 observed photons compared against ${4\!\times\!10^{4}}$ simulated photons. 

A Kolmogorov--Smirnov test finds a maximum difference between the two
cumulative radial distributions, ${\text{D}\!=\!0.525}$ (see Fig\,\ref{map}).
This implies that the source associated to ULX1 is not pointlike with a 
confidence level of $5\sigma$ (${\text{p-value}\!=\!3\!\times\!10^{-7}}$).
We repeated the same analysis simulating extended sources with a disk profile
of uniform surface brightness. 
We can reject at the $2\sigma$ confidence level all values of the disk radius
outside the range ${R_{\text{d}}\!=\!1.35\pm0.50\text{\,arcsec}}$.
We repeated the same analysis assuming a smooth halo shape\cite{draine03}
with a hole, as described in the section of the Supplementary Material where
we apply a dust scattering model.
We constrained the values of the distance between the source and a dust layer
to ${d_\text{sd}\!=\!11 \pm 5 \text{\,kpc}}$.
In the spirit of reproducible results, we provide the code of the MARX plugin
(Draine\_halo.c) that we used to simulate such a dust halo at
\href{https://github.com/andrea-belfiore/MARX-plugins.git}{https://github.com/andrea-belfiore/MARX-plugins.git}.

\begin{figure}
%\internallinenumbers % required by nature
\centering
\resizebox{\hsize}{!}{\includegraphics{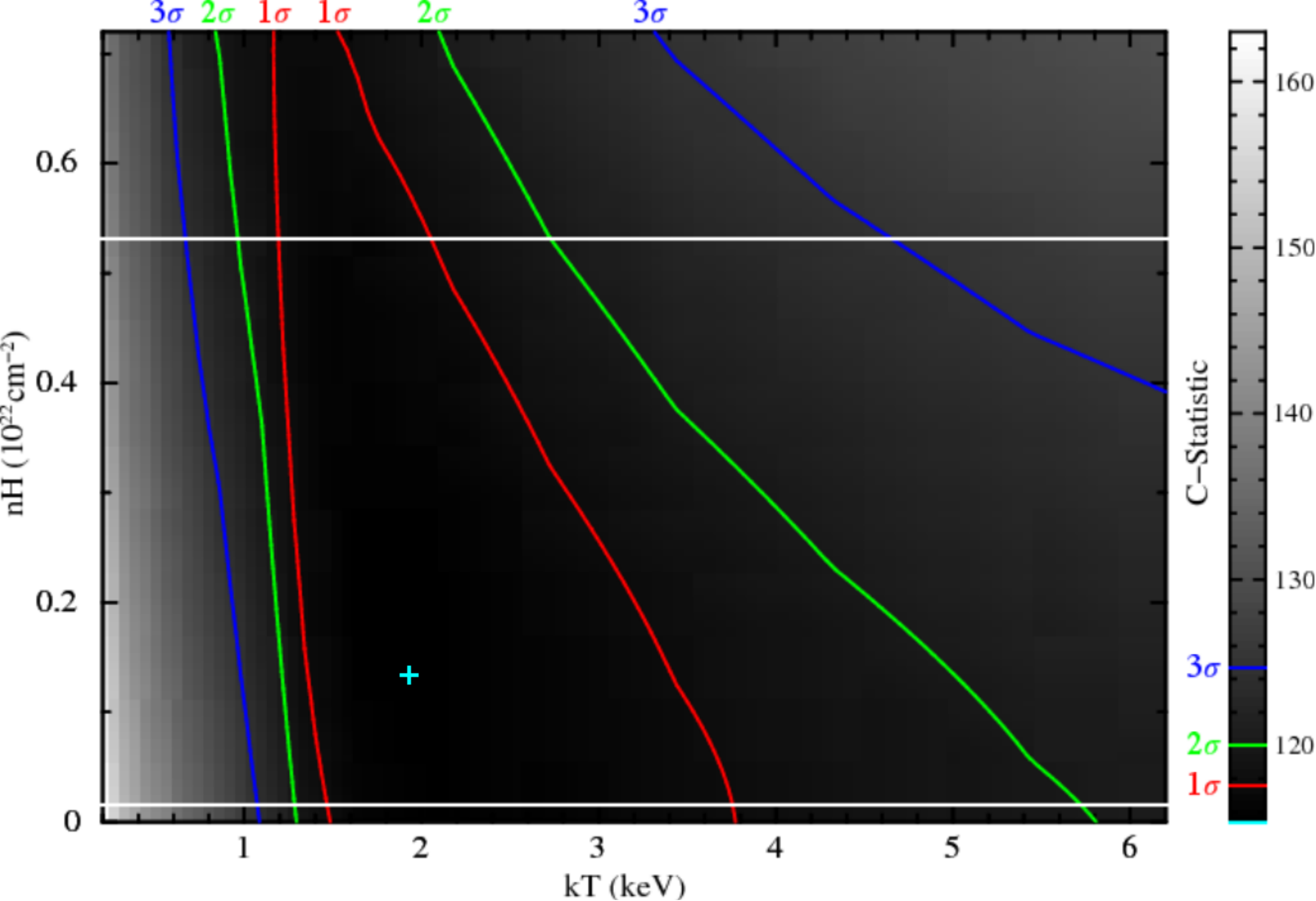}}
\caption{
\textbf{Joint spectral fit to the plasma temperature and the absorption column
obtained with Xspec and an apec model, of the extended emission observed
with \cxo\ in November 2017 (Obs.~Id.~20830).}
The cyan cross corresponds to the best fit on these X-ray data.
The red, green, and blue color contours indicate the 1, 2, and 3$\sigma$ 
confidence levels, respectively.
Although the column density is poorly constrained by these X-ray data,
we have two external constraints, marked as horizontal white lines in
this plot.
The lower value, $N_{\mathrm{H,G}} = 1.2 \times 10^{20} \text{cm}^{-2}$, is given
by the Galactic absorption in the direction of ULX1.
The higher value $N_{\mathrm{H,U}} = 5.3 \times 10^{21} \text{cm}^{-2}$, was estimated
in the high state of ULX1, and thus might include internal absorption.
}
\label{cxo_spectrum}
\end{figure}

%%%%%%%%%%%%%%%%%%%%%%%%%%%%%%%%%%%%%%%%%%%%%%%%%%%%%%%%%%%%%%%%%%%%%%%%%%%%%%%
\noindent\emph{Spectral analysis}\\
We analysed the spectrum of the extended source with XSPEC, adopting
a collisional plasma model (apec), absorbed according to the T{\"u}bingen-Boulder 
absorption model (tbabs) and set the abundances to those of ref.\cite{wilms00}.
As the number of events is very low, we used C-statistics\cite{cash79} and
verified through Monte Carlo simulations similar to those described above,
that we obtain identical estimates and error bars.
All the uncertainties are stated at 90\% CL.
The absorption column is poorly constrained by the data (see 
Fig.\,\ref{cxo_spectrum}), but physical constraints are given by the absorption
level of our Galaxy in the direction of  ULX1, 
$N_{H,G} = 1.2 \times 10^{20} \text{cm}^{-2}$, and the absorption measured
in the high state of ULX1, $N_{H,U} = 5.3 \times 10^{21} \text{cm}^{-2}$
(that includes internal absorption by the ULX itself).
We repeated the same analysis adopting the best-fit value
$N_{H} = 1.3 \times 10^{21} \text{cm}^{-2}$, that provided us with the
best-fit estimates for all parameters, and for the two extreme values
$N_{H,G}$ and $N_{H,U}$ that determine our uncertainties (that cover the
90\% CL error bars for all allowed values of $N_{H}$).

%%%%%%%%%%%%%%%%%%%%%%%%%%%%%%%%%%%%%%%%%%%%%%%%%%%%%%%%%%%%%%%%%%%%%%%%%%%%%%%

%%%%%%%%%%%%%%%%%%%%%%%%%%%%%%%%%%%%%%%%%%%%%%%%%%%%%%%%%%%%%%%%%%%%%%%%%%%%%%%
We estimate a characteristic energy $\text{k}_B T=1.9_{-0.8}^{+2.3}\text{ keV}$,
a normalisation ${N=3.5_{-1.9}^{+2.5} \times 10^{-6}}$, and an unabsorbed
0.3-7.0 \text{\,keV} flux 
${F_X=6.1_{-2.5}^{+6.1}\times10^{-15}\text{\,\flux}}$,
corresponding to an isotropic 0.3-7.0 \text{\,keV} luminosity
${L_{\mathrm{X}} = 2.1_{-0.9}^{+2.1}\!\times\!10^{38}\text{\,\lum}}$.

We also tested through Monte Carlo simulations a dust-scattering spectrum
expected from the model outlined in a following section. 
This model is acceptable at the $2\sigma$ level (the spectrum changes so
slightly with the tuning parameters that we cannot constrain them in this
way).

We try to estimate the flux of a point-like component associated to ULX1.
To this aim, we consider only the innermost photons, within 
${0.5\text{\,arcsec}}$ from the location of ULX1, determined as described
in the astrometry section. 
We assume for the point source a power-law spectrum with ${\Gamma\!=\!2}$
and the absorption column $N_{H,U} = 5.3 \times 10^{21} \text{cm}^{-2}$
measured for ULX1 in its high state.
We used the CIAO tool srcflux, that applies an aperture photometry, taking
into account the response of the instrument, the encircled fraction (fraction
of the PSF within each  region), statistical uncertainties, position on the
detector, and other effects. It provides an upper limit on the flux (at the
90\% confidence level) of
${F_{\text{X}}\!<\!3.4\!\times\!10^{-15}\text{\,\flux}}$.
This corresponds to a luminosity
${L_{\text{X}}\!<\!1.2\!\times\!10^{38}\text{\,\lum}\!=\!0.68\,L_{14}}$, 
where ${L_{14}\!=\!1.764\!\times\!10^{38}\text{\,\lum}}$\ is the Eddington 
limit for a 1.4-$M_{\odot}$ neutron star.

%%%%%%%%%%%%%%%%%%%%%%%%%%%%%%%%%%%%%%%%%%%%%%%%%%%%%%%%%%%%%%%%%%%%%%%%%%%%%%%
\noindent\textbf{Physical modeling of the expanding nebula}\\
We assume that an isotropic wind with constant power ${L_{\text{w}}}$ 
is emitted from the ULX system, shocks the external medium (ISM) and expands 
according to a self-similar solution.\cite{castor75,weaver77} 
After a short free expansion period, the wind forms a shock that starts as 
adiabatic but becomes more and more radiatively efficient. 
In this phase, the expanding nebula (bubble) is radially structured in 4
regions:
\vspace*{-.5cm}
\begin{enumerate}
  \item Close to the source is the low-density left-over of the swept up ISM,
      where the wind expands freely (free wind region);
  \item Starting at a radius ${R_{1}}$ the wind, faster than the shock, 
      accumulates, increasing density and temperature (shocked wind region);
  \item Starting at a radius $R_{\text{c}}$ the swept up ISM accumulates, 
      increasing density and temperature (shocked ISM region);
  \item Beyond a radius $R_{2}$ the ISM is still unperturbed by the shock
      (ISM region).
\end{enumerate}
\vspace*{-.5cm}
In this model, there are two shocked regions where X-rays could be emitted: 
the shocked wind region and the shocked ISM region. 
The first scenario was considered in a set of simulations\cite{siwek17} of 
expanding nebulae, and, with the boundary conditions set in these simulations,
it cannot reach an X-ray luminosity larger than $\sim$$ 10^{35}\text{\,\lum}$.
Either this scenario is missing some dominant effect, or it can be ruled out
in the context of ULX1; we explore here the other scenario.

If we assume that the X-rays are produced in the shocked ISM region, then 
the observed disk radius ${R_{\text{d}}\!=\!1.35\pm0.50\text{\,arcsec}}$
coincides with ${R_{2}( \tau )\!=\!112\pm42\text{\,pc}}$ at a distance
${D\!=\!17.1\text{\,Mpc}}$, where $\tau$ is the current age of the bubble.
Strong shock conditions for the temperature of the shocked ISM provide
a direct estimate of the current speed of the shock $v_\text{sh}$:
\begin{equation}
  v_{\text{sh}}^{(0)} = \sqrt{ 
      \frac{16}{3} \text{\,} \frac{kT}{m_{\text{p}}}
      } = \left( 3.3_{-0.8}^{+1.6} \right) \times 10^{-3} c = 
      \left( 9.9_{-2.4}^{+4.8} \right) \times 10^{2} \text{\,km\,s}^{-1}
      \label{eq:v_sh}
\end{equation}
Because observations of other nebulae show that the shock speed
is somewhat overestimated with this approximation, we introduce
a factor $\xi<1$ that accounts for this discrepancy:
$v_{\text{sh}} = v_{\text{sh}}^{(0)} \times \xi$.
A standard bubble model, with a constant injection of mechanical power
${L_{\text{w}}}$ in a uniform ISM with density 
$n_{\text{ISM}}\!=\!n_{1}\!\times\!\text{cm}^{-3}$, predicts a time 
dependence of the bubble size:
\begin{equation}
  R_{2}(t) = \alpha \left(\frac{L_{\text{w}} 
  t^{3} }{ n_{\text{ISM}}m_{\text{p}}} \right)
  ^{\nicefrac{1}{5}}\label{eq:R2}
\end{equation}
where $\alpha\!=\!0.88$ is a numerical constant.\cite{weaver77}

%%%%%%%%%%%%%%%%%%%%%%%%%%%%%%%%%%%%%%%%%%%%%%%%%%%%%%%%%%%%%%%%%%%%%%%%%%%%%%%
Under these assumptions, as 
$v_{\text{sh}}\!=\!\frac{\text{d}R_{2}}{\text{d}t}$, we can estimate
the age of the bubble:
\begin{equation}
  \tau = \frac{3}{5} \frac{R_{2}}{v_{\text{sh}}} 
      = \left( 6.7_{-2.8}^{+3.1} \right) \times10^{4}\text{\,yr}\times \xi ^{-1}
      \label{eq:tau}
\end{equation}

%%%%%%%%%%%%%%%%%%%%%%%%%%%%%%%%%%%%%%%%%%%%%%%%%%%%%%%%%%%%%%%%%%%%%%%%%%%%%%%
As ${v_{\text{sh}}}$ largely exceeds the sound speed in the ISM, the
Rankine--Hugoniot conditions at the shock front grant that the plasma density
just inside the shocked region is 
${n_{\text{sh}} = W \times n_{\text{ISM}}}$, with a compression factor
$W \simeq 4$.
However, as a high radiative efficiency strongly increases this
value, we maintain it as a free parameter.
According to the standard bubble model, ${n_{\text{sh}}}$ decreases
down to 0 as the radial distance $r$ approaches a contact discontinuity
at ${r\!=\!R_{\text{c}}}$.
Because the bremsstrahlung emissivity 
${\epsilon\!\propto\!n_{\text{sh}}^2}$, we assume that most
emission is produced close to ${R_{2}}$.
We estimate the size of this emitting region $V_{\text{sh}}$ by
assuming that it contains most of the swept up material from the ISM:
\begin{equation}
  V_{\text{sh}} \simeq \frac{4}{3} \pi R_{2}^{3} 
      \frac{n_{\text{ISM}}}{n_{\text{sh}}} = 
      \left( 1.5_{-1.1}^{+2.3} \right) \times10^{6}\text{\,pc}^{3}
      \times \left( \frac{W}{4} \right) ^{-1}
\end{equation}

We can now extract from the normalisation of the apec model an estimate
of the shocked plasma density as 
\begin{equation}
    n_{\mathrm{sh}} = 3.1_{-2.1}^{+11}\times10^{-1}\text{ cm}^{-3} \times
        \left(\frac{W}{4}\right)^{\frac{1}{2}}
\end{equation}
This value is broadly consistent with those observed in optical/radio ULX
bubbles.\!\cite{pakull10}

%%%%%%%%%%%%%%%%%%%%%%%%%%%%%%%%%%%%%%%%%%%%%%%%%%%%%%%%%%%%%%%%%%%%%%%%%%%%%%%
We use eq.\,\eqref{eq:R2} to provide an estimate of the mechanical power
of the wind $L_{\text{w}}$:
\begin{equation}
  L_{\mathrm{w}} = \frac{m_{\text{p}} n_\text{ISM} R_2^5 }{\alpha^5 \tau^3} \simeq
    108 m_{\text{p}} n_\text{ISM} R_2^2
    \left(\frac{k_B T}{m_{\text{p}}}\right)^{\!\nicefrac{3}{2}} \!\!\!\times \xi^{3}
    \!\!= \left(1.3_{-1.0}^{+9.8}\right) \times 10^{41}\text{\,\lum} \!\times\!
    \xi^{3} \!\times\! \left(\frac{W}{4}\right)^{\!-\frac{1}{2}}
    \label{eq:L_wind}
\end{equation}
where we used eqq.\,\eqref{eq:tau} and \eqref{eq:v_sh} and our estimates for
${R_\text{d}}$, $k_B T$, and ${n_{\text{ISM}}}$, leading to large
uncertainties.
This value is in the same order of magnitude as the X-ray luminosity observed
from ULX1 at its peak, if emitted isotropically.\!\cite{israel17}
Although $L_{\text{w}}$ is weakly constrained from our results, we deem it
extremely unlikely that it can be much higher.

Indeed, if we suppose that ULX1 sustained this value of $L_{\text{w}}$ for
the age of the nebula $\tau$, its wind would have carried an energy 
${E\!=\!L_{\text{w}}\tau =
    \left(2.8_{-2.2}^{+18}\right) \times 10^{53}\text{ erg} \times
    \xi^{2} \times \left(\frac{W}{4}\right)^{\!-\frac{1}{2}}}$.
As the wind, and therefore the bubble, is powered by accretion onto a
neutron star, the accreted mass must be at least:
\begin{equation}
  M_\text{accr} = \frac{E}{\eta c^2} \simeq 
      \left(1.8^{+12}_{-1.5}\right) \times 10^{33} \text{\,g} = 0.9^{+5.9}_{-0.7} \,M_\odot
\end{equation}
where $\eta \simeq \frac{GM_{ns}}{R_{ns}c^2} \simeq 17\%$ is the
accretion efficiency of a neutron star with mass $M_{ns}\!=\!1.4 M_{\odot}$
and radius $R_{ns}\!=\!12\text{\,km}$. 
The best-fit value of ${M_\text{accr}}$ might be too large for a neutron 
star, as it would have probably already collapsed into a black hole, but
a value of ${M_\text{accr} < 0.5\,M_\odot}$ is more plausible.
As the accretion power is not all channelled into the wind, but must also
sustain the luminosity of ULX1, it seems likely that our simple model
might need some revision.

%%%%%%%%%%%%%%%%%%%%%%%%%%%%%%%%%%%%%%%%%%%%%%%%%%%%%%%%%%%%%%%%%%%%%%%%%%%%%%%
We can relate the total mass ejected by the wind, ${M_\text{w}}$ to $E$ and,
indirectly, to ${M_\text{accr}}$:
\begin{equation}
  M_\text{w} = \frac{2 \eta M_\text{accr} c^2}{v_{\text{w}}^2}
\end{equation}
where $v_{\text{w}}$ is the speed of the wind.
If we assume $M_\text{accr} \simeq 0.5\text{\,}M_\odot$, then, to keep 
$M_\text{w} \apprle 10 M_\odot$ as expected for an X-ray binary system,
$v_{\text{w}} > 0.1\text{\,}c$. 
This value of $v_{\text{w}}$ agrees with the velocity measured in outflows
from various ULXs.\cite{pinto16,kosec18,walton16b}

\clearpage

\part*{Supplementary Information}

\noindent\textbf{Detectability of similar features}\\
We explore the conditions under which an extended feature associated to
a point source, particularly a ULX, is detectable.
Due to the small angular scale of such a feature, we restrict our scope
to \cxo, which provides the best angular resolution among the currently
operating X-ray observatories.
We simulate with MARX a large number of extended sources overlapping a point
source and a uniform background, scanning a broad range for each parameter: 
the size (disk radius $R_{\mathrm{d}}$) and the fluence of the extended source, the 
fluence of the point source ($N_{\mathrm{p}}$) and that of the background.
We assume for the extended source a uniform disk shape and a bremsstrahlung
spectrum with ${kT\!=\!0.76\,\text{keV}}$, and for the point source a power-law 
spectrum with index ${\Gamma\!=\!2}$.

By scanning over the fluence of the extended source, we determine for each
configuration of the other parameters the number of photons $N_{90}$ that
are needed from the extended source to detect it as extended with a 90\%
probability.
We consider a source detected as extended if we can reject at the $3\sigma$
confidence level the hypothesis that the source is point-like, if we apply
a procedure identical to the one we applied to the source associated to ULX1.
It turns out that the background level has an appreciable impact only in those
cases where the number of events from the two source components (one point-like,
the other extended) combined is very small.
Therefore, we fix the number of expected background photons within
$3\,\text{arcsec}$ to $4$, close to the value we estimate in the field of ULX1.
We also find that fixing $N_{\mathrm{p}}$ and increasing $R_{\mathrm{d}}$, $N_{90}$ 
decreases down to a minimum value for $R_{\mathrm{d}}\gtrsim 2\,\text{arcsec}$.

%%%%%%%%%%%%%%%%%%%%%%%%%%%%%%%%%%%%%%%%%%%%%%%%%%%%%%%%%%%%%%%%%%%%%%%%%%%%%%%
\begin{figure}
%\internallinenumbers % required by nature
\centering
\resizebox{\hsize}{!}{\includegraphics{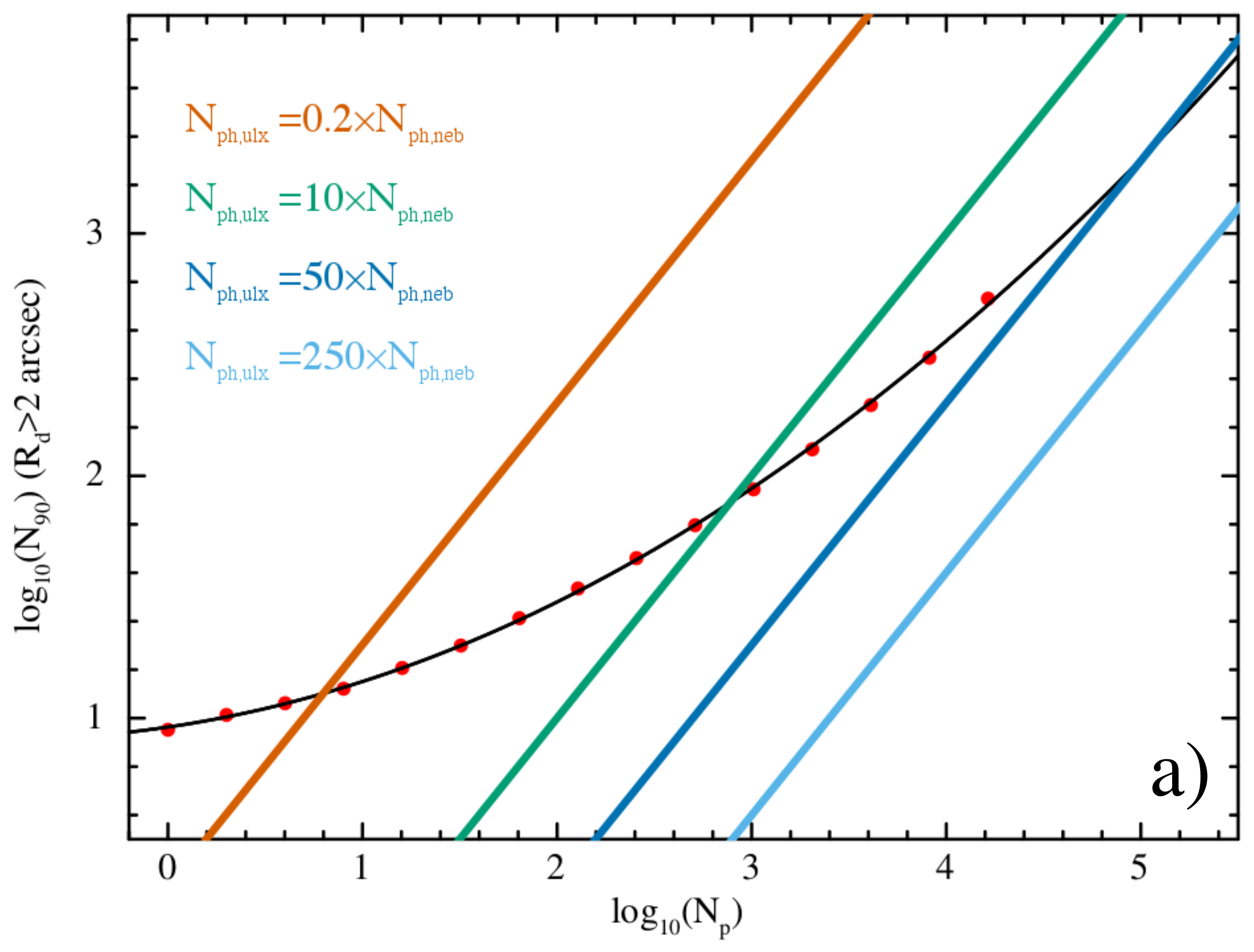} \includegraphics{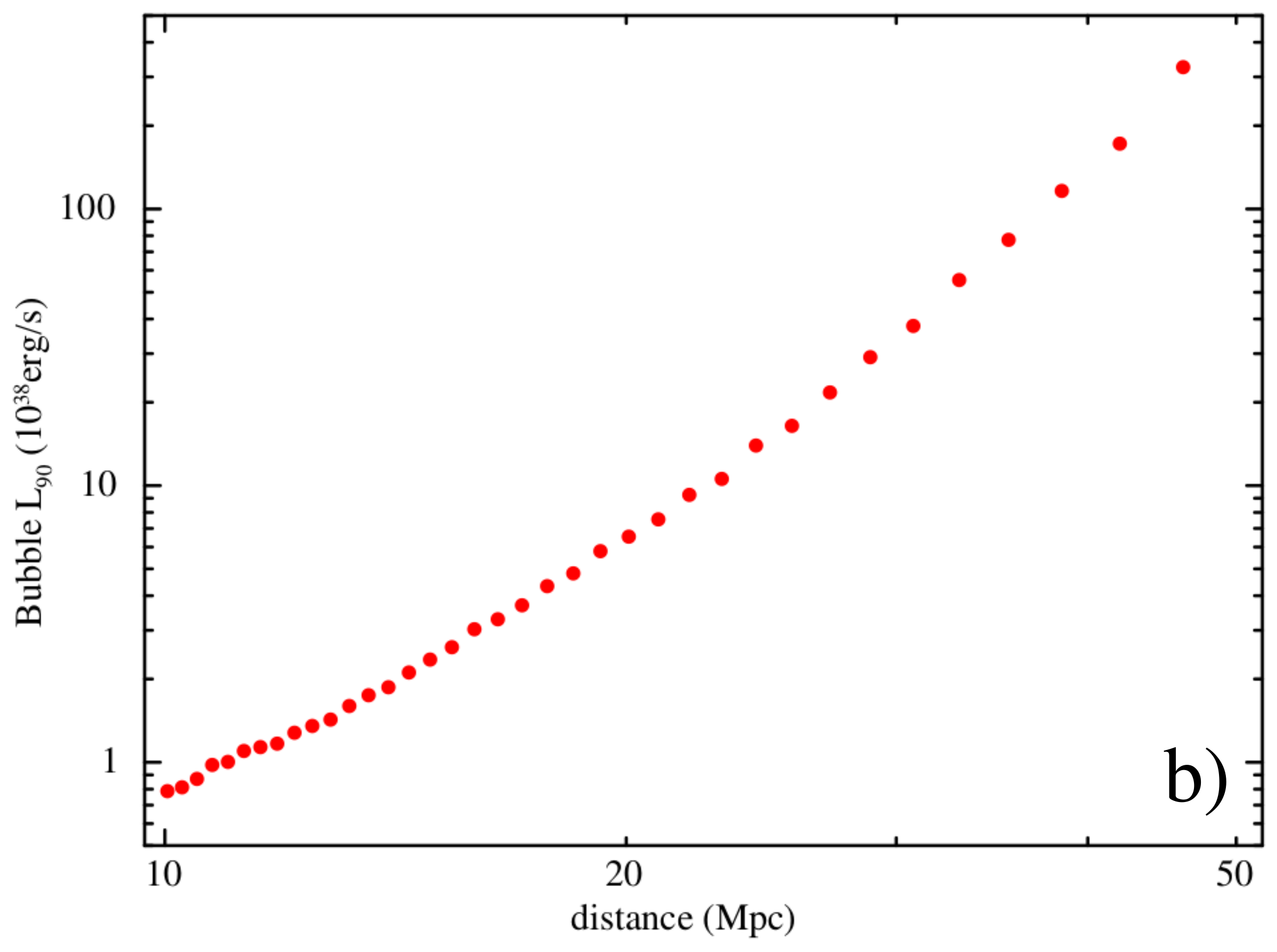}}
\caption{
\textbf{Detectability of a ULX nebula with \cxo.}
\textbf{a)}: Minimum number of photons from a sufficiently extended
source ($R_{\mathrm{d}}\!\gtrsim\!2\,\text{arcsec}$), on top of a point source, 
for it to appear extended. 
This value grants a 90\% probability of rejecting at the $3\sigma$ confidence level
the hypothesis that the source is point-like.
The axes report the base-10 logarithms of counts in the point-like (x-axis) and
extended (y-axis) components.
Straight lines indicate a fixed ratio of counts between the two components.
Only if the point-like component is much dimmer than the extended component,
the extended feature can be detected.
\textbf{b)}: Minimum luminosity of an extended feature to be
detected as a function of the distance of its host galaxy.
The spectra of the components, their relative flux ratio, and physical size
of the extended component are fixed to the best-fit values for \src.
A feature dimmer than ${7\!\times\!10^{39}\text{\,\lum}}$ can be detected only
within $20\,\text{Mpc}$.
}
\label{detectability}
\end{figure}

The left hand panel of Suppl.\,Fig.\,\ref{detectability} shows $N_{90}$ as a 
function of $N_{\mathrm{p}}$ for a relatively large source, with 
${R_{\mathrm{d}}\gtrsim 2\,\text{arcsec}}$.
For a feature similar in physical size as the one we detect around ULX1, 
this corresponds to a distance of the host galaxy ${\lesssim\,12\,\text{Mpc}}$.
We overplot a few lines indicating fixed count ratios between the point
source and the extended source.
If we assume that the brightness of the extended feature is constant,
the blue and cyan lines correspond to the counts ratios that we expect when
ULX1 is in its high state, the green line the counts ratio in the
intermediate state, and the red line the counts ratio in the low state.
The left hand panel of Suppl.\,Fig.\,\ref{detectability} shows that a ULX must be
in a low state for us to be able to detect a dim extended feature associated
to it, even if the source is not too far from us.

Then, we assume that the extended source is 5 times more luminous than the
point source (reproducing the off state of a transient ULX) and check its
detectability at larger distance $\gtrsim\,10\,\text{Mpc}$.
We consider a $50\,\text{ks}$ \cxo\ observation and the same physical size
for the extended feature as the one we detect around ULX1.
The right hand panel of Suppl.\,Fig.\,\ref{detectability} shows the luminosity of
the extended feature ($L_{90}$) corresponding to $N_{90}$ as a function
of the distance of the host galaxy.
If we assume that the X-ray luminosity of the nebula should be lower than
a few ${10^{39}\text{\,\lum}}$, then we can only appreciate the extension
of sources within $\sim$$20\,\text{Mpc}$ (and only when the point source
is in a very low state).
Very few currently known ULXs are eligible,\!\cite{walton11} and they should
be closely monitored, because only when they are in a low state \cxo\ could 
probe with a long stare if they share a feature similar to the one we observe
in ULX1 in this paper.

%%%%%%%%%%%%%%%%%%%%%%%%%%%%%%%%%%%%%%%%%%%%%%%%%%%%%%%%%%%%%%%%%%%%%%%%%%%%%%%
\noindent\textbf{Application of a dust scattering model}\label{dust}\\
We investigated the possibility that the extended emission around ULX1 is due
to X-ray dust-scattering, as sometimes it is observed for X-ray sources whose
emission passes through clouds of dust (e.g.
refs.\,\cite{predehl95,vaughan04,vasilopoulos16}).
The observed angular deviation $\phi$ is related\cite{mathis91}
to the scattering angle $\theta$ and to the source and dust distances
($D$ and $D_{\rm d}$, respectively) by  ${\phi \simeq (1 - x) \times \theta}$,
where $x=D_{\rm d}/D$. For dust located in our Galaxy and not too
close to the X-ray source (since the X-ray scattering cross section
rapidly drops with the scattering angle), we can see scattering
halos around X-ray sources up to dozens of arcmin. 
An extended X-ray scattering halo by extragalactic dust, instead, has never
been detected because for $x\!\approx\!1$, the uncertainty in the PSF
hampers a measure of the tiny extent of these halos.
However, the favourable combination of the \cxo\ PSF, the edge-on
orientation of  NGC\,5907, the extreme luminosity of ULX1, and its abrupt
switch-off, might have given us the opportunity to resolve
spatially---for the first time---the X-ray emission from an extragalactic
compact object that was scattered by the dust in its host galaxy.

In fact, we can interpret the halo we saw as due to the photons emitted
during the high state of ULX1 and scattered in our direction by the
interstellar dust in its host-galaxy, the difference between the two optical
paths causing a delay. 
Indeed, dust in NGC\,5907 prevents us to see any optical counterpart to ULX1,
even with the high sensitivity of the Hubble Space Telescope.
In the following section, we apply the X-ray scattering model to 
our system and check its consistency.

%%%%%%%%%%%%%%%%%%%%%%%%%%%%%%%%%%%%%%%%%%%%%%%%%%%%%%%%%%%%%%%%%%%%%%%%%%%%%%%
As a first approximation, we assume that the dust is uniformly distributed
in a thin wall at a distance 
${d_{\mathrm{sd}} = (1 - x) \times D = d_{10} \times 10\text{\,kpc}}$ from ULX1,
where  ${D=17.1\text{\,Mpc}}$ is the distance of NGC\,5907.
The difference in optical path with respect to a photon reaching us directly
causes a delay:
\begin{equation}
  \Delta t\left( x, \phi \right) \simeq
      \frac{d_{\mathrm{sd}} \left((1-x)\theta^{2}+\phi^{2}\right)}{2c (1-x)}
      \simeq \frac{D \phi^{2}}{2c (1-x)} \label{eq:Dt}.
\end{equation}
The brightness profile of the halo can be expressed\cite{mathis91} as:
\begin{equation}
  B_{\text{h}}\left( x, t, \phi \right) =  N_{\mathrm{d}}
      L_{\mathrm{src}}\left( t - \Delta t\left( x, \phi \right) \right)
      \int S(E) \frac{\text{d}\sigma}{\text{d}\theta}
      \left( \frac{\phi}{1-x}, E \right) \text{\,d}E, \label{eq:B_halo}
\end{equation}
where $N_{\mathrm{d}}$ is the dust (scattering centers) column density,
$L_{\mathrm{src}}$ is the X-ray luminosity of the source, $S(E)$ is the
spectral energy distribution of the source normalized to 1, and
$\frac{\text{d}\sigma}{\text{d}\theta}$ is the differential scattering
cross-section.

%%%%%%%%%%%%%%%%%%%%%%%%%%%%%%%%%%%%%%%%%%%%%%%%%%%%%%%%%%%%%%%%%%%%%%%%%%%%%%%
If $L_{\mathrm{src}}$ switches on at an epoch $t_{\mathrm{on}}$, then, according to
eqq.\,\eqref{eq:B_halo} and \eqref{eq:Dt}, a halo forms up to a radius:
\begin{equation}
  \phi_{\mathrm{max}}(x, t) \simeq \sqrt{\frac{2c (1 - x)}{D} 
      \left(t-t_{\mathrm{on}}\right)} \label{eq:phi_max}
\end{equation}
If $L_{\mathrm{src}}$ has kept constant over a long time, then
$B_{\mathrm{h}}\left(\phi \right)$ fades out at large $\phi$ because
$\frac{\text{d}\sigma}{\text{d}\theta}$ drops at large $\theta$.
Therefore, while $\frac{\text{d}\sigma}{\text{d}\theta}$ shapes the
radial brightness profile of the halo, $x$ (or $d_\mathrm{sd}$, if $D$ is 
known) fully determines the halo size.
Although a precise expression for $\frac{\text{d}\sigma}{\text{d}\theta}$
depends on the dust properties, we can approximate its 90\% containment 
angle\cite{draine03} as ${\theta_{90} \simeq 0.3^\circ \times E_{1}^{-1}}$
where $E=E_{1}\text{\,keV}$ is the photon energy.
Since the lowest energy photons determine the size of the halo and 21 out
of the 25 photons we used for our extension analysis have an energy
${E>\text{1\,keV}}$, we assume that ${\theta_{90} \simeq 0.3^\circ}$.
After a time 
${t - t_\mathrm{on} \simeq \frac{D}{2c} (1-x) \theta_{90}^{2}
\simeq 160 \text{\,d} \times d_{10} }$
the halo saturates to an angular size:
\begin{equation}
  \phi_{\mathrm{max}}(x, E) \simeq (1 - x) \theta_{90}(E)
      \simeq 0.63\text{\,arcsec}\times d_{10}
\end{equation}
As with the activation of the source, if $L_{\mathrm{src}}$ abruptly drops to 0 
at an epoch $t_\mathrm{off}$, then
${B_{\mathrm{h}}\left(\phi\!<\!\phi_{\mathrm{min}} \right)\!=\!0}$ with:
\begin{equation}
  \phi_{\mathrm{min}}(x, t) \simeq \sqrt{\frac{2c (1 - x)}{D} 
    \left(t-t_{\mathrm{off}}\right)} \label{eq:phi_min}
\end{equation}
In this simple model we expect a sharp hole expanding in the halo starting
from its center, without altering the profile of the halo for 
${\phi > \phi_{\mathrm{min}}}$.
In a more realistic case, in which the source takes some time to switch
off and light scatters in a smooth distribution of dust at various $x$, 
we expect a more complex halo profile and evolution.

%%%%%%%%%%%%%%%%%%%%%%%%%%%%%%%%%%%%%%%%%%%%%%%%%%%%%%%%%%%%%%%%%%%%%%%%%%%%%%%
% INTERNAL NOTE ON THE DUST SCATTERING CROSS SECTION (MARIO)
% Draine 2013 provides an analytic approximation to the cross section
% for dust scattering. His rule of thumb for the 90% containment radius
% is R_90 = 0.3 deg / (E/keV).
% After some thought I am convinced that this value should be compared against
% the 2D 90% containment radius in a Gaussian fit (R_90(2D) = 2.303 sigma)
% However, these values (and the distance that follows) heavily rely on the 
% dust model. Tiengo et al 2010 show a comparison of results for different 
% models with a broad range of variability. Their distance estimate using 
% Draine's (WD01) model is 6.91 kpc, BARE-GR_B gives the smallest distance
% 3.91 kpc and COMP-NC-B the largest, 11.83 kpc. Therefore, we can introduce
% an uncertainty by a factor 2 in the final constraints on d_{sd} to account
% for the dust properties.
%%%%%%%%%%%%%%%%%%%%%%%%%%%%%%%%%%%%%%%%%%%%%%%%%%%%%%%%%%%%%%%%%%%%%%%%%%%%%%%
As \swift\ monitoring shows that ULX1 remained in a high state
(${L_{\mathrm{on}} \simeq 10^{41} \text{\,\lum}}$) for $\sim$4
years\cite{walton16,israel17}, before switching off, we expect that the halo
is complete, for any reasonable value of $d_{\mathrm{sd}}$ (up to $\sim$90 kpc).
We observed ULX1 switching off on 2017 July 10 (our $t_{\mathrm{off}}$)
and \cxo\ observation 20830 took place on 2017 November 7, 120\,d later.
Therefore, we expect a hole in the halo profile with a radius
${\phi_{\mathrm{min}} \simeq 0.54\text{\,arcsec}
    \times d_{10}^{\nicefrac{1}{2}}}$.
As described in the extension analysis section, our fit to a halo profile
estimates at the 90\% confidence level that
${d_{\mathrm{sd}}\!=\!11 \pm 5\text{\,kpc}}$.

\begin{figure}
%\internallinenumbers % required by nature
\centering
\resizebox{.97\hsize}{!}{\includegraphics{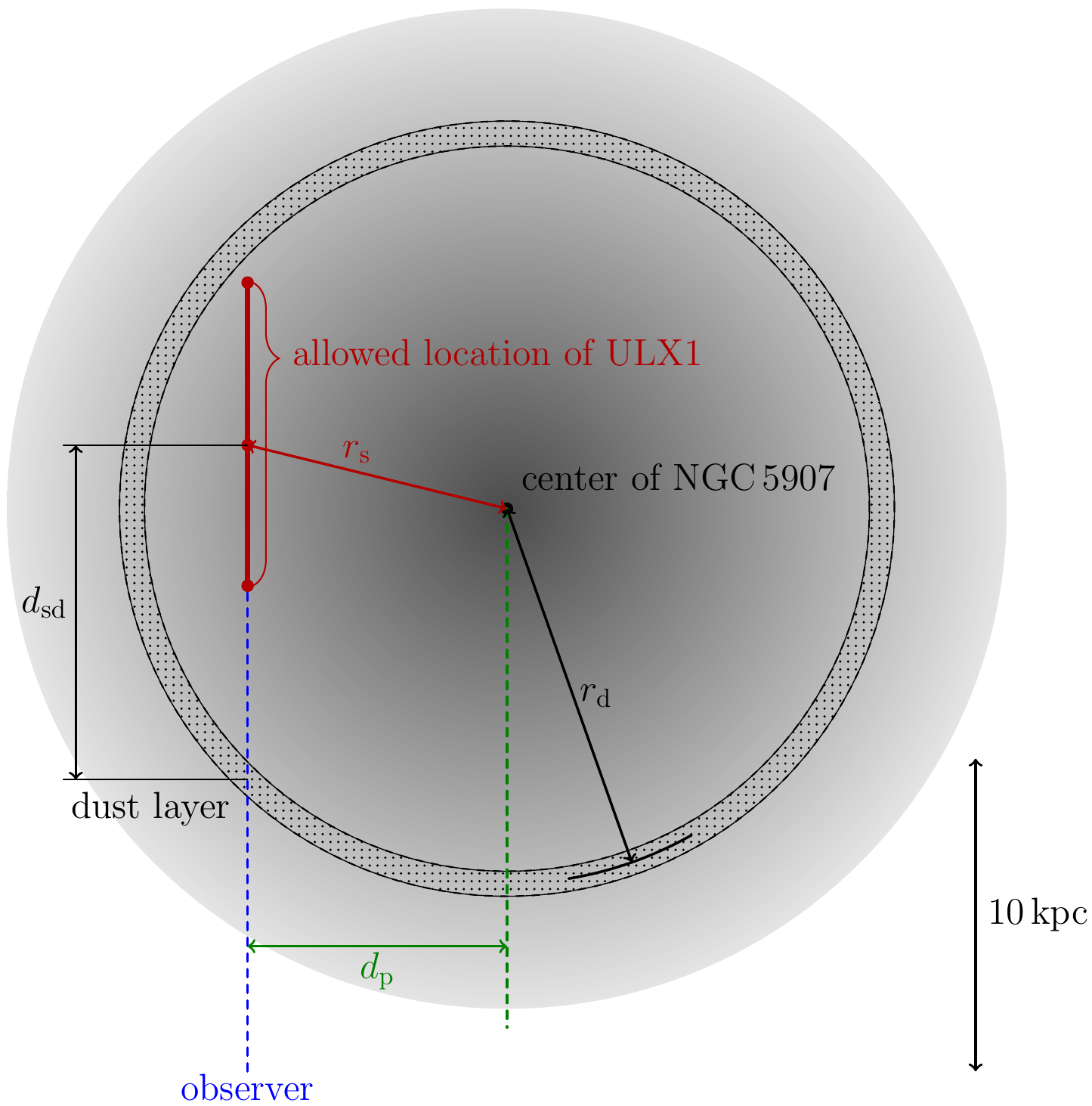}}
\caption{
\textbf{Schematic representation of NGC\,5907 viewed face-on, and
geometrical constraints provided by the dust-scattering scenario.}
The shaded grey region reproduces the radial dust distribution observed 
in NGC\,5907.\cite{xilouris99}
If we assume that a dust wall is located at $r_{\mathrm{d}}=12$\,kpc from
the center of NGC\,5907, then all distances are in scale, apart for the
shape of the dust layer, here pictured as a thick ring, but whose global
structure we ignore.
The projected distance of ULX1 from the center of NGC\,5907 is
${d_{\mathrm{p}}=8.3\text{\,kpc}}$.
The distance between the dust layer and ULX1 is constrained by the model
to ${d_{\mathrm{sd}}=11 \pm 5\text{\,kpc}}$.
The thick vertical red line, along the blue dashed line of sight, marks
the range of physical locations of ULX1 allowed by this model.
The distance between ULX1 and the center of NGC\,5907 is $r_{\mathrm{s}}$.
As the line of sight crosses a sizable portion of the plane of NGC\,5907,
optical and X-ray absorption should be very large.}
\label{toy}
\end{figure}

%%%%%%%%%%%%%%%%%%%%%%%%%%%%%%%%%%%%%%%%%%%%%%%%%%%%%%%%%%%%%%%%%%%%%%%%%%%%%%%
% INTERNAL NOTE ON THE GEOMETRY OF NGC 5907 (MARIO)
% I believe Xilouris et al 1999 provide the most accurate and robust model
% of the galaxy 3D structure to date. They find an inclination of 87.2+/-2 deg.
% They assume a distance of 11 Mpc, though, so their geometrical estimates
% must be rescaled accordingly (boosted by a factor 1.555).
% They fit a 3D double exponential model, radially symmetric.
% For dust they report: h_d=5.30(13) kpc and z_d=0.113(6) kpc
% which is rescaled to H_d=8.23(20) kpc and Z_d=0.176(9) kpc
% For stars they report: h_s=4.60(6) kpc and z_s=0.333(6) kpc
% which is rescaled to H_s=7.15(9) kpc and Z_s=0.518(9) kpc
%%%%%%%%%%%%%%%%%%%%%%%%%%%%%%%%%%%%%%%%%%%%%%%%%%%%%%%%%%%%%%%%%%%%%%%%%%%%%%%
We consider now the structure and geometry of NGC\,5907 to constrain
the location of ULX1 within its host galaxy.
A model that accounts for optical emission and extinction\cite{xilouris99}
estimates the inclination of the galaxy plane as
${87.\!\!^\circ2 \pm 0.\!\!^\circ2}$.
A double exponential spatial model sets the scale length for dust and stars
as ${h_{\mathrm{d}}\!=\!8.2 \pm 0.2\text{\,kpc}}$, and
${h_{\mathrm{s}}\!=\!7.2 \pm 0.1\text{\,kpc}}$, respectively.
The dust radial profile shows a steep decrease at $\sim$16\,kpc, while 
the gas extends much further out\cite{alton04}.
We approximate NGC\,5907 to a perfectly thin edge-on galaxy and
apply a simple geometrical model sketched, face-on, in Suppl.\,Fig.\,\ref{toy}.
The distance between ULX1 and a dust layer is ${d_{\mathrm{sd}}}$,
where $r_{\mathrm{d}}$ is the distance of the dust cloud from the center
of NGC\,5907, $r_{\mathrm{s}}$ is the distance of ULX1 from the center
of NGC\,5907, and ${d_{\mathrm{p}}\!=\!\text{8.3\,kpc}}$ is its sky projection.
Given the constraints above, the optical depth is minimal when
${r_{\mathrm{d}}\!=\!16\text{\,kpc}}$ and ${d_{\mathrm{sd}}\!=\!6\text{\,kpc}}$.

%%%%%%%%%%%%%%%%%%%%%%%%%%%%%%%%%%%%%%%%%%%%%%%%%%%%%%%%%%%%%%%%%%%%%%%%%%%%%%%

We follow a standard approach that assumes that the dust distribution,
and therefore optical absorption, is a good proxy to track the metals
that cause the X-ray absorption.\!\cite{watson11}
The numerical integration of a model describing the optical absorption 
in NGC\,5907\cite{just06}, along the line of sight up to ULX1, leads to
${6.4\text{\,mag}\!<\!A_{\mathrm{V}}\!<\!12.6\text{\,mag}}$.
Assuming Solar abundances, we convert\cite{watson11} these values
into an estimate of the hydrogen column density: 
${1.4\!\times\!10^{22}\text{\,cm}^{-2}\!<\!N_{\mathrm{H}}
\!<\!2.8\!\times\!10^{22}\text{\,cm}^{-2}}$.
This estimate is not consistent with the value of 
${N_{\mathrm{H}}\!=\!5.3\!\times\!10^{21}\text{\,cm}^{-2}}$ 
we see from ULX1 in its high state\cite{israel17} (which includes absorption
internal to the source).
Uncertainties in the distribution and properties of the dust prevent
us from drawing firm conclusions but, since the dust-scattering
hypothesis seems to require some degree of fine tuning in the 
dust and galaxy parameters to reproduce the data, we do not 
further explore this scenario.

%%%%%%%%%%%%%%%%%%%%%%%%%%%%%%%%%%%%%%%%%%%%%%%%%%%%%%%%%%%%%%%%%%%%%%%%%%%%%%%

\noindent\textbf{Multi-wavelength Coverage}\\
The region around ULX1 has been observed in $H_{\alpha}$ with the 0.9 m
telescope at the Kitt Peak National Observatory\cite{rand96} and in radio
at 5 GHz with the Very Large Array in configuration B\cite{mezcua13}.
The $H_{\alpha}$ image of NGC 5907 shows a large number of excesses on
the order of a few $10^{-15}\text{\,\flux}$, that follow the profile
of the galaxy. 
One of them, with a flux ${F_{\alpha} \approx 9\times10^{-16}\text{\flux}}$,
is consistent with the direction of ULX1.
However, the poor angular resolution ($\sim 1\text{\,arcsec}$) hampers a firm
association with the X-ray nebula, or any further characterisation
of the $H_{\alpha}$ feature.
The radio observation reveals no point source down to a flux density of
$20\,\mu \text{Jy}$ within 10 arcsec around the direction of ULX1.

For a fully radiative strong collisional shock we expect that the flux in the
Balmer lines is proportional to the mechanical power\cite{weaver77}
${L_{\mathrm{m}}=\frac{12}{55} L_{\mathrm{w}}}$ and depends
on the shock speed.\cite{dopita96,allen08} 
Most of the recombination happens in the photoionized pre-shock region, or
in the post-shock region where the plasma has already cooled.
Because the age of the nebula is smaller than the cooling time of the plasma,
we expect this contribution to be negligible.

The MAPPINGS III shock model library 
tables (\url{http://cdsweb.u-strasbg.fr/~allen/mappings_page1.html})
including both the pre- and post-shock regions, lead to an expected
unabsorbed flux in $H_{\beta}$:
\[
  F_{\beta} =    
      \left(1.53 \pm 0.20\right) \times 10^{-16} \text{\,\flux} \times
      \frac{L_{w}}{10^{40}\text{\,\lum}} \times
      d_{17}^{-\frac{1}{2}}
\]
and an expected unabsorbed flux in $H_{\alpha}$:
\[
  F_{\alpha} =   
      \left(4.7 \pm 0.7\right) \times 10^{-16} \text{\,\flux} \times
      \frac{L_{w}}{10^{40}\text{\,\lum}} \times
      d_{17}^{-\frac{1}{2}}
\]
This value, obtained under the assumption of a fully radiative strong shock,
is in slight tension with the excess observed by the Kitt Peak National
Observatory, if we adopt the estimate of $L_w$ obtained in
the adiabatic assumption, eq.6 in the main text.
However, the strong optical absorption in the direction of
ULX1\cite{heida19} might reduce this tension.

% both versions
The bremsstrahlung emissivity of a hot plasma at radio frequencies can be
written\cite{rybicki86} as:
\[
j_{\nu}=\frac{8e^{6}}{3mc^{2}}\left(\frac{2\pi}{3mc^{2}}\right)^{\frac{1}{2}}\left(kT\right)^{-\frac{1}{2}}Z^{2}n_{i}n_{e}\frac{\sqrt{3}}{\pi}\ln\left(\frac{4}{\zeta}\frac{kT}{h\nu}\right)
\]
where $\zeta=1.781...$ is the exponential Euler-Mascheroni constant.
For pure hydrogen, with a compression factor $W\gtrsim4$, we expect a radio
flux density:
\[
F_{5\text{GHz}}=1.74\times10^{-3}\text{ \ensuremath{\mu}Jy}\times\left(\frac{W}{4}\right)\times\left(\frac{n_{\mathrm{ism}}}{\text{cm}^{-3}}\right)^{2}\times\left(\frac{R}{112\text{ pc}}\right)^{3}\times\left(\frac{kT}{\text{keV}}\right)^{-\frac{1}{2}}\times\left(18.5+\ln\left(\frac{T}{\text{keV}}\right)\right)
\]
We expect no detectable radio emission from the hot X-ray-emitting plasma.
Instead, under the assumption of a fully radiative strong shock, we expect some
detectable radio emission from other regions.

If we combine the above estimate for $F_{\beta}$ to the relation between
the $H_{\beta}$ and the $5\text{ GHz}$ emissivities\cite{caplan86} 
for a plasma with $k_{\mathrm{B}}T=1.9\text{ keV}$, we obtain:
\[
  F_{5\text{GHz}} = 12.0\text{ }\mu\text{Jy} \times
      \frac{F_{\beta}}{10^{-16}\text{ erg cm}^{-2}\text{s}^{-1}}
      = 18.3 \pm 2.4 \text{\,}\mu\text{Jy} \times
      \frac{L_{\mathrm{w}}}{10^{40}\text{\,\lum}} \times
      d_{17}^{-\frac{1}{2}}
\]
% both versions
VLA did observe at 5GHz NGC 5907 ULX-1 for 3h on May 30 2012, in the
B configuration,\cite{mezcua13}
reporting an upper limit of $20\text{ }\mu\text{Jy beam}^{-1}$.
Because the beam width at half power for VLA at 5GHz in configuration B
measures $1.2 \text{ arcsec}$, we expect that the bubble may be covered
by a few beams, depending on the size of the radio nebula.
Deriving a limit on the brightness of extended sources with an aperture
synthesis array is a non-trivial task,\cite{crane89} that goes beyond
the scope of this paper.

The tension between our estimates and the reported upper limit
indicates that a fully radiative shock approximation might not apply
to our case. Indeed, such an approximation implies that all the
mechanical power $L_{\mathrm{m}}$ is radiated, while, according to our 
model of the nebula,
${L_{\mathrm{bol}} \approx L_{\mathrm{X}} \ll L_{\mathrm{m}}}$.
% both version
Dropping the assumption of a fully radiative shock requires a more
sophisticated and complete model of the nebula, that accounts for multiple
plasma phases.

\begin{addendum}

%%%%%%%%%%%%%%%%%%%%%%%%%%%%%%%%%%%%%%%%%%%%%%%%%%%%%%%%%%%%%%%%%%%%%%%%%%%%%%%
\item
This research is based on observations made with the Chandra X-ray Observatory
and has made use of software provided by the \cxo\ X-ray Center (CXC) in the 
application packages CIAO, ChIPS, and Sherpa.
This research also made use of data obtained with the Neil Gehrels Swift
Observatory and \xmm.
\swift\ is a NASA mission with participation of the Italian Space Agency and
the UK Space Agency.
\xmm\ is an ESA science mission with instruments and contributions directly
funded by ESA Member States and NASA. 
A.B. is grateful to A. Fabian for an interesting discussion and to S. Covino
for help in optical data reduction.
A.B. and G.N. are supported by EXTraS, a project funded by the European
Union’s (EU) Seventh Framework Programme under grant agreement no. 607452.
We acknowledge funding in the framework of the project ULTraS (ASI--INAF
contract N.\,2017-14-H.0).
M.M. acknowledges funding from ASI--INAF contract N.\,2015-023-R.0.
D.J.W. acknowledges financial support from an STFC Ernest Rutherford
Fellowship.

\end{addendum}


\newpage

\noindent{\large\bf References}
\begin{thebibliography}{10}
\expandafter\ifx\csname url\endcsname\relax
  \def\url#1{\texttt{#1}}\fi
\expandafter\ifx\csname urlprefix\endcsname\relax\def\urlprefix{URL }\fi
\providecommand{\bibinfo}[2]{#2}
\providecommand{\eprint}[2][]{\url{#2}}
\bibitem{feng11}
\bibinfo{author}{{Feng}, H.} \& \bibinfo{author}{{Soria}, R.}
\newblock \bibinfo{title}{{Ultraluminous X-ray sources in the Chandra and
  XMM-Newton era}}.
\newblock \emph{\bibinfo{journal}{\nar}} \textbf{\bibinfo{volume}{55}},
  \bibinfo{pages}{166--183} (\bibinfo{year}{2011}).

\bibitem{kaaret17}
\bibinfo{author}{{Kaaret}, P.}, \bibinfo{author}{{Feng}, H.} \&
  \bibinfo{author}{{Roberts}, T.~P.}
\newblock \bibinfo{title}{{Ultraluminous X-Ray Sources}}.
\newblock \emph{\bibinfo{journal}{\araa}} \textbf{\bibinfo{volume}{55}},
  \bibinfo{pages}{303--341} (\bibinfo{year}{2017}).

\bibitem{bachetti14}
\bibinfo{author}{{Bachetti}, M.} \emph{et~al.}
\newblock \bibinfo{title}{{An ultraluminous X-ray source powered by an
  accreting neutron star}}.
\newblock \emph{\bibinfo{journal}{\nat}} \textbf{\bibinfo{volume}{514}},
  \bibinfo{pages}{202--204} (\bibinfo{year}{2014}).

\bibitem{furst16}
\bibinfo{author}{{F{\"u}rst}, F.} \emph{et~al.}
\newblock \bibinfo{title}{{Discovery of Coherent Pulsations from the
  Ultraluminous X-Ray Source NGC 7793 P13}}.
\newblock \emph{\bibinfo{journal}{\apjl}} \textbf{\bibinfo{volume}{831}},
  \bibinfo{pages}{L14} (\bibinfo{year}{2016}).

\bibitem{israel17}
\bibinfo{author}{{Israel}, G.~L.} \emph{et~al.}
\newblock \bibinfo{title}{{An accreting pulsar with extreme properties drives
  an ultraluminous x-ray source in NGC 5907}}.
\newblock \emph{\bibinfo{journal}{Science}} \textbf{\bibinfo{volume}{355}},
  \bibinfo{pages}{817--819} (\bibinfo{year}{2017}).

\bibitem{ipe17}
\bibinfo{author}{{Israel}, G.~L.} \emph{et~al.}
\newblock \bibinfo{title}{{Discovery of a 0.42-s pulsar in the ultraluminous
  X-ray source NGC 7793 P13}}.
\newblock \emph{\bibinfo{journal}{\mnras}} \textbf{\bibinfo{volume}{466}},
  \bibinfo{pages}{L48--L52} (\bibinfo{year}{2017}).

\bibitem{carpano18}
\bibinfo{author}{{Carpano}, S.}, \bibinfo{author}{{Haberl}, F.},
  \bibinfo{author}{{Maitra}, C.} \& \bibinfo{author}{{Vasilopoulos}, G.}
\newblock \bibinfo{title}{{Discovery of pulsations from NGC 300 ULX1 and its
  fast period evolution}}.
\newblock \emph{\bibinfo{journal}{\mnras}} \textbf{\bibinfo{volume}{476}},
  \bibinfo{pages}{L45--L49} (\bibinfo{year}{2018}).

\bibitem{drury12}
\bibinfo{author}{{Drury}, L.~O.~.}
\newblock \bibinfo{title}{{Origin of cosmic rays}}.
\newblock \emph{\bibinfo{journal}{Astroparticle Physics}}
  \textbf{\bibinfo{volume}{39}}, \bibinfo{pages}{52--60}
  (\bibinfo{year}{2012}).

\bibitem{abeysekara18}
\bibinfo{author}{{Abeysekara}, A.~U.} \emph{et~al.}
\newblock \bibinfo{title}{Very-high-energy particle acceleration powered by the
  jets of the microquasar SS 433}.
\newblock \emph{\bibinfo{journal}{\nat}} \textbf{\bibinfo{volume}{562}},
  \bibinfo{pages}{82--85} (\bibinfo{year}{2018}).

\bibitem{xilouris99}
\bibinfo{author}{{Xilouris}, E.~M.}, \bibinfo{author}{{Byun}, Y.~I.},
  \bibinfo{author}{{Kylafis}, N.~D.}, \bibinfo{author}{{Paleologou}, E.~V.} \&
  \bibinfo{author}{{Papamastorakis}, J.}
\newblock \bibinfo{title}{{Are spiral galaxies optically thin or thick?}}
\newblock \emph{\bibinfo{journal}{\aap}} \textbf{\bibinfo{volume}{344}},
  \bibinfo{pages}{868--878} (\bibinfo{year}{1999}).

\bibitem{tully13}
\bibinfo{author}{{Tully}, R.~B.} \emph{et~al.}
\newblock \bibinfo{title}{{Cosmicflows-2: The Data}}.
\newblock \emph{\bibinfo{journal}{\aj}} \textbf{\bibinfo{volume}{146}},
  \bibinfo{pages}{86} (\bibinfo{year}{2013}).

\bibitem{just06}
\bibinfo{author}{{Just}, A.}, \bibinfo{author}{{M{\"o}llenhoff}, C.} \&
  \bibinfo{author}{{Borch}, A.}
\newblock \bibinfo{title}{{An evolutionary disc model of the edge-on galaxy NGC
  5907}}.
\newblock \emph{\bibinfo{journal}{\aap}} \textbf{\bibinfo{volume}{459}},
  \bibinfo{pages}{703--716} (\bibinfo{year}{2006}).

\bibitem{blondin98}
\bibinfo{author}{{Blondin}, J.~M.}, \bibinfo{author}{{Wright}, E.~B.},
  \bibinfo{author}{{Borkowski}, K.~J.} \& \bibinfo{author}{{Reynolds}, S.~P.}
\newblock \bibinfo{title}{{Transition to the Radiative Phase in Supernova
  Remnants}}.
\newblock \emph{\bibinfo{journal}{\apj}} \textbf{\bibinfo{volume}{500}},
  \bibinfo{pages}{342--354} (\bibinfo{year}{1998}).

\bibitem{fryer12}
\bibinfo{author}{{Fryer}, C.~L.} \emph{et~al.}
\newblock \bibinfo{title}{{Compact Remnant Mass Function: Dependence on the
  Explosion Mechanism and Metallicity}}.
\newblock \emph{\bibinfo{journal}{\apj}} \textbf{\bibinfo{volume}{749}},
  \bibinfo{pages}{91} (\bibinfo{year}{2012}).

\bibitem{barkov11}
\bibinfo{author}{{Barkov}, M.~V.} \& \bibinfo{author}{{Komissarov}, S.~S.}
\newblock \bibinfo{title}{{Recycling of neutron stars in common envelopes and
  hypernova explosions}}.
\newblock \emph{\bibinfo{journal}{\mnras}} \textbf{\bibinfo{volume}{415}},
  \bibinfo{pages}{944--958} (\bibinfo{year}{2011}).
\newblock \eprint{1012.4565}.

\bibitem{pakull02}
\bibinfo{author}{{Pakull}, M.~W.} \& \bibinfo{author}{{Mirioni}, L.}
\newblock \bibinfo{title}{{Optical Counterparts of Ultraluminous X-Ray
  Sources}}.
\newblock In \emph{\bibinfo{booktitle}{New Visions of the X-ray Universe in the
  XMM-Newton and Chandra Era, 26-30 November 2001, ESTEC, The Netherlands}},
  ArXiv Astrophysics e-prints: astro-ph/0202488 (\bibinfo{year}{2002}).

\bibitem{pakull03}
\bibinfo{author}{{Pakull}, M.~W.} \& \bibinfo{author}{{Mirioni}, L.}
\newblock \bibinfo{title}{{Bubble Nebulae around Ultraluminous X-Ray Sources}}.
\newblock In \bibinfo{editor}{{Arthur}, J.} \& \bibinfo{editor}{{Henney},
  W.~J.} (eds.) \emph{\bibinfo{booktitle}{Revista Mexicana de Astronomia y
  Astrofisica Conference Series}}, vol.~\bibinfo{volume}{15} of
  \emph{\bibinfo{series}{Revista Mexicana de Astronomia y Astrofisica,
  vol.~27}}, \bibinfo{pages}{197--199} (\bibinfo{year}{2003}).

\bibitem{pakull08}
\bibinfo{author}{{Pakull}, M.~W.} \& \bibinfo{author}{{Gris{\'e}}, F.}
\newblock \bibinfo{title}{{Ultraluminous X-ray Sources: Beambags and Optical
  Counterparts}}.
\newblock In \bibinfo{editor}{{Bandyopadhyay}, R.~M.},
  \bibinfo{editor}{{Wachter}, S.}, \bibinfo{editor}{{Gelino}, D.} \&
  \bibinfo{editor}{{Gelino}, C.~R.} (eds.) \emph{\bibinfo{booktitle}{A
  Population Explosion: The Nature \& Evolution of X-ray Binaries in Diverse
  Environments}}, vol. \bibinfo{volume}{1010} of
  \emph{\bibinfo{series}{American Institute of Physics Conference Series}},
  \bibinfo{pages}{303--307} (\bibinfo{year}{2008}).

\bibitem{lang07}
\bibinfo{author}{{Lang}, C.~C.}, \bibinfo{author}{{Kaaret}, P.},
  \bibinfo{author}{{Corbel}, S.} \& \bibinfo{author}{{Mercer}, A.}
\newblock \bibinfo{title}{{A Radio Nebula Surrounding the Ultraluminous X-Ray
  Source in NGC 5408}}.
\newblock \emph{\bibinfo{journal}{\apj}} \textbf{\bibinfo{volume}{666}},
  \bibinfo{pages}{79--85} (\bibinfo{year}{2007}).

\bibitem{kaaret03}
\bibinfo{author}{{Kaaret}, P.}, \bibinfo{author}{{Corbel}, S.},
  \bibinfo{author}{{Prestwich}, A.~H.} \& \bibinfo{author}{{Zezas}, A.}
\newblock \bibinfo{title}{{Radio Emission from an Ultraluminous X-ray Source}}.
\newblock \emph{\bibinfo{journal}{Science}} \textbf{\bibinfo{volume}{299}},
  \bibinfo{pages}{365--368} (\bibinfo{year}{2003}).

\bibitem{castor75}
\bibinfo{author}{{Castor}, J.}, \bibinfo{author}{{McCray}, R.} \&
  \bibinfo{author}{{Weaver}, R.}
\newblock \bibinfo{title}{{Interstellar bubbles}}.
\newblock \emph{\bibinfo{journal}{\apjl}} \textbf{\bibinfo{volume}{200}},
  \bibinfo{pages}{L107--L110} (\bibinfo{year}{1975}).

\bibitem{weaver77}
\bibinfo{author}{{Weaver}, R.}, \bibinfo{author}{{McCray}, R.},
  \bibinfo{author}{{Castor}, J.}, \bibinfo{author}{{Shapiro}, P.} \&
  \bibinfo{author}{{Moore}, R.}
\newblock \bibinfo{title}{{Interstellar bubbles. II - Structure and
  evolution}}.
\newblock \emph{\bibinfo{journal}{\apj}} \textbf{\bibinfo{volume}{218}},
  \bibinfo{pages}{377--395} (\bibinfo{year}{1977}).

\bibitem{pakull10}
\bibinfo{author}{{Pakull}, M.~W.}, \bibinfo{author}{{Soria}, R.} \&
  \bibinfo{author}{{Motch}, C.}
\newblock \bibinfo{title}{{A 300-parsec-long jet-inflated bubble around a
  powerful microquasar in the galaxy NGC 7793}}.
\newblock \emph{\bibinfo{journal}{\nat}} \textbf{\bibinfo{volume}{466}},
  \bibinfo{pages}{209--212} (\bibinfo{year}{2010}).

\bibitem{dopita12}
\bibinfo{author}{{Dopita}, M.~A.}, \bibinfo{author}{{Payne}, J.~L.},
  \bibinfo{author}{{Filipovi{\'c}}, M.~D.} \& \bibinfo{author}{{Pannuti},
  T.~G.}
\newblock \bibinfo{title}{{The physical parameters of the microquasar S26 in
  the Sculptor Group galaxy NGC 7793}}.
\newblock \emph{\bibinfo{journal}{\mnras}} \textbf{\bibinfo{volume}{427}},
  \bibinfo{pages}{956--967} (\bibinfo{year}{2012}).

\bibitem{pinto16}
\bibinfo{author}{{Pinto}, C.}, \bibinfo{author}{{Middleton}, M.~J.} \&
  \bibinfo{author}{{Fabian}, A.~C.}
\newblock \bibinfo{title}{{Resolved atomic lines reveal outflows in two
  ultraluminous X-ray sources}}.
\newblock \emph{\bibinfo{journal}{\nat}} \textbf{\bibinfo{volume}{533}},
  \bibinfo{pages}{64--67} (\bibinfo{year}{2016}).

\bibitem{kosec18}
\bibinfo{author}{{Kosec}, P.} \emph{et~al.}
\newblock \bibinfo{title}{{Evidence for a variable Ultrafast Outflow in the
  newly discovered Ultraluminous Pulsar NGC 300 ULX-1}}.
\newblock \emph{\bibinfo{journal}{\mnras}} \textbf{\bibinfo{volume}{479}},
  \bibinfo{pages}{3978--3986} (\bibinfo{year}{2018}).

\bibitem{walton16b}
\bibinfo{author}{{Walton}, D.~J.} \emph{et~al.}
\newblock \bibinfo{title}{{An Iron K Component to the Ultrafast Outflow in NGC
  1313 X-1}}.
\newblock \emph{\bibinfo{journal}{\apjl}} \textbf{\bibinfo{volume}{826}},
  \bibinfo{pages}{L26} (\bibinfo{year}{2016}).

\bibitem{rand96}
\bibinfo{author}{{Rand}, R.~J.}
\newblock \bibinfo{title}{{Diffuse Ionized Gas in Nine Edge-on Galaxies}}.
\newblock \emph{\bibinfo{journal}{\apj}} \textbf{\bibinfo{volume}{462}},
  \bibinfo{pages}{712} (\bibinfo{year}{1996}).

\bibitem{mezcua13}
\bibinfo{author}{{Mezcua}, M.}, \bibinfo{author}{{Roberts}, T.~P.},
  \bibinfo{author}{{Sutton}, A.~D.} \& \bibinfo{author}{{Lobanov}, A.~P.}
\newblock \bibinfo{title}{{Radio observations of extreme ULXs: revealing the
  most powerful ULX radio nebula ever or the jet of an intermediate-mass black
  hole?}}
\newblock \emph{\bibinfo{journal}{\mnras}} \textbf{\bibinfo{volume}{436}},
  \bibinfo{pages}{3128--3134} (\bibinfo{year}{2013}).

\bibitem{dopita96}
\bibinfo{author}{{Dopita}, M.~A.} \& \bibinfo{author}{{Sutherland}, R.~S.}
\newblock \bibinfo{title}{{Spectral Signatures of Fast Shocks. I. Low-Density
  Model Grid}}.
\newblock \emph{\bibinfo{journal}{\apjs}} \textbf{\bibinfo{volume}{102}},
  \bibinfo{pages}{161} (\bibinfo{year}{1996}).

\bibitem{caplan86}
\bibinfo{author}{{Caplan}, J.} \& \bibinfo{author}{{Deharveng}, L.}
\newblock \bibinfo{title}{{Extinction and reddening of H II regions in the
  Large Magellanic Cloud.}}
\newblock \emph{\bibinfo{journal}{\aap}} \textbf{\bibinfo{volume}{155}},
  \bibinfo{pages}{297--313} (\bibinfo{year}{1986}).

\bibitem{begelman06}
\bibinfo{author}{{Begelman}, M.~C.}, \bibinfo{author}{{King}, A.~R.} \&
  \bibinfo{author}{{Pringle}, J.~E.}
\newblock \bibinfo{title}{{The nature of SS433 and the ultraluminous X-ray
  sources}}.
\newblock \emph{\bibinfo{journal}{\mnras}} \textbf{\bibinfo{volume}{370}},
  \bibinfo{pages}{399--404} (\bibinfo{year}{2006}).

\bibitem{walton16}
\bibinfo{author}{{Walton}, D.~J.} \emph{et~al.}
\newblock \bibinfo{title}{{A 78 Day X-Ray Period Detected from NGC 5907 ULX1 by
  Swift}}.
\newblock \emph{\bibinfo{journal}{\apjl}} \textbf{\bibinfo{volume}{827}},
  \bibinfo{pages}{L13} (\bibinfo{year}{2016}).

\end{thebibliography}

\noindent{\large\bf References}
\begin{thebibliography}{10}
\expandafter\ifx\csname url\endcsname\relax
  \def\url#1{\texttt{#1}}\fi
\expandafter\ifx\csname urlprefix\endcsname\relax\def\urlprefix{URL }\fi
\providecommand{\bibinfo}[2]{#2}
\providecommand{\eprint}[2][]{\url{#2}}

\makeatletter
\addtocounter{\@listctr}{33}
\makeatother

\bibitem{garmire03}
\bibinfo{author}{{Garmire}, G.~P.}, \bibinfo{author}{{Bautz}, M.~W.},
  \bibinfo{author}{{Ford}, P.~G.}, \bibinfo{author}{{Nousek}, J.~A.} \&
  \bibinfo{author}{{Ricker}, G.~R., Jr.}
\newblock \bibinfo{title}{{Advanced CCD imaging spectrometer (ACIS) instrument
  on the Chandra X-ray Observatory}}.
\newblock In \bibinfo{editor}{{Truemper}, J.~E.} \&
  \bibinfo{editor}{{Tananbaum}, H.~D.} (eds.) \emph{\bibinfo{booktitle}{X-Ray
  and Gamma-Ray Telescopes and Instruments for Astronomy.}}, vol.
  \bibinfo{volume}{4851} of \emph{\bibinfo{series}{Proceedings of the SPIE.}},
  \bibinfo{pages}{28--44} (\bibinfo{publisher}{SPIE, Bellingham},
  \bibinfo{year}{2003}).

\bibitem{fruscione06}
\bibinfo{author}{{Fruscione}, A.} \emph{et~al.}
\newblock \bibinfo{title}{{CIAO: Chandra's data analysis system}}.
\newblock In \bibinfo{editor}{{Silva}, D.~R.} \& \bibinfo{editor}{{Doxsey},
  R.~E.} (eds.) \emph{\bibinfo{booktitle}{Observatory Operations: Strategies,
  Processes, and Systems}}, vol. \bibinfo{volume}{6270} of
  \emph{\bibinfo{series}{SPIE Conference Series}}, \bibinfo{pages}{62701V}
  (\bibinfo{publisher}{SPIE, Bellingham}, \bibinfo{year}{2006}).

\bibitem{struder01}
\bibinfo{author}{{Str{\"u}der}, L.} \emph{et~al.}
\newblock \bibinfo{title}{{The European Photon Imaging Camera on XMM-Newton:
  The pn-CCD camera}}.
\newblock \emph{\bibinfo{journal}{\aap}} \textbf{\bibinfo{volume}{365}},
  \bibinfo{pages}{L18--L26} (\bibinfo{year}{2001}).

\bibitem{turner01}
\bibinfo{author}{{Turner}, M.~J.~L.} \emph{et~al.}
\newblock \bibinfo{title}{{The European Photon Imaging Camera on XMM-Newton:
  The MOS cameras}}.
\newblock \emph{\bibinfo{journal}{\aap}} \textbf{\bibinfo{volume}{365}},
  \bibinfo{pages}{L27--L35} (\bibinfo{year}{2001}).

\bibitem{gabriel04}
\bibinfo{author}{{Gabriel}, C.} \emph{et~al.}
\newblock \bibinfo{title}{{The XMM-Newton SAS - Distributed Development and
  Maintenance of a Large Science Analysis System: A Critical Analysis}}.
\newblock In \bibinfo{editor}{{Ochsenbein}, F.}, \bibinfo{editor}{{Allen},
  M.~G.} \& \bibinfo{editor}{{Egret}, D.} (eds.)
  \emph{\bibinfo{booktitle}{Astronomical Data Analysis Software and Systems
  (ADASS) XIII}}, vol. \bibinfo{volume}{314} of \emph{\bibinfo{series}{ASP
  Conf. Ser. (San Francisco, CA: ASP)}}, \bibinfo{pages}{759}
  (\bibinfo{year}{2004}).

\bibitem{pintore18}
\bibinfo{author}{{Pintore}, F.} \emph{et~al.}
\newblock \bibinfo{title}{{A new ultraluminous X-ray source in the galaxy NGC
  5907}}.
\newblock \emph{\bibinfo{journal}{\mnras}} \textbf{\bibinfo{volume}{477}},
  \bibinfo{pages}{L90--L95} (\bibinfo{year}{2018}).

\bibitem{arnaud96}
\bibinfo{author}{{Arnaud}, K.~A.}
\newblock \bibinfo{title}{{XSPEC: The First Ten Years}}.
\newblock In \bibinfo{editor}{{Jacoby}, G.~H.} \& \bibinfo{editor}{{Barnes},
  J.} (eds.) \emph{\bibinfo{booktitle}{Astronomical Data Analysis Software and
  Systems V}}, vol. \bibinfo{volume}{101} of
  \emph{\bibinfo{series}{\it Astron.~Soc.~Pac.~Conf.~Series}},
  \bibinfo{pages}{17--20} (\bibinfo{year}{1996}).

\bibitem{wilms00}
\bibinfo{author}{{Wilms}, J.}, \bibinfo{author}{{Allen}, A.} \&
  \bibinfo{author}{{McCray}, R.}
\newblock \bibinfo{title}{{On the Absorption of X-Rays in the Interstellar
  Medium}}.
\newblock \emph{\bibinfo{journal}{\apj}} \textbf{\bibinfo{volume}{542}},
  \bibinfo{pages}{914--924} (\bibinfo{year}{2000}).

\bibitem{burrows05}
\bibinfo{author}{{Burrows}, D.~N.} \emph{et~al.}
\newblock \bibinfo{title}{{The Swift X-Ray Telescope}}.
\newblock \emph{\bibinfo{journal}{Space Science Reviews}}
  \textbf{\bibinfo{volume}{120}}, \bibinfo{pages}{165--195}
  (\bibinfo{year}{2005}).

\bibitem{blackburn95}
\bibinfo{author}{{Blackburn}, J.~K.}
\newblock \bibinfo{title}{{FTOOLS: A FITS Data Processing and Analysis Software
  Package}}.
\newblock In \emph{\bibinfo{booktitle}{{Shaw}, R.~A. and {Payne}, H.~E. and
  {Hayes}, J.~J.~E., eds., Astronomical Data Analysis Software and Systems
  IV.}}, vol.~\bibinfo{volume}{77} of \emph{\bibinfo{series}{\it Astron.~Soc.~Pac.~Conf.~Series}}, \bibinfo{pages}{367} (\bibinfo{year}{1995}).

\bibitem{davis12}
\bibinfo{author}{{Davis}, J.~E.} \emph{et~al.}
\newblock \bibinfo{title}{{Raytracing with MARX: x-ray observatory design,
  calibration, and support}}.
\newblock In \emph{\bibinfo{booktitle}{Space Telescopes and Instrumentation
  2012: Ultraviolet to Gamma Ray}}, vol. \bibinfo{volume}{8443} of
  \emph{\bibinfo{series}{\procspie}}, \bibinfo{pages}{84431A}
  (\bibinfo{publisher}{SPIE, Bellingham}, \bibinfo{year}{2012}).

\bibitem{draine03}
\bibinfo{author}{{Draine}, B.~T.}
\newblock \bibinfo{title}{{Scattering by Interstellar Dust Grains. II.
  X-Rays}}.
\newblock \emph{\bibinfo{journal}{\apj}} \textbf{\bibinfo{volume}{598}},
  \bibinfo{pages}{1026--1037} (\bibinfo{year}{2003}).

\bibitem{cash79}
\bibinfo{author}{{Cash}, W.}
\newblock \bibinfo{title}{{Parameter estimation in astronomy through
  application of the likelihood ratio}}.
\newblock \emph{\bibinfo{journal}{\apj}} \textbf{\bibinfo{volume}{228}},
  \bibinfo{pages}{939--947} (\bibinfo{year}{1979}).

\bibitem{siwek17}
\bibinfo{author}{{Siwek}, M.}, \bibinfo{author}{{S{\c a}dowski}, A.},
  \bibinfo{author}{{Narayan}, R.}, \bibinfo{author}{{Roberts}, T.~P.} \&
  \bibinfo{author}{{Soria}, R.}
\newblock \bibinfo{title}{{Optical and X-ray luminosities of expanding nebulae
  around ultraluminous X-ray sources}}.
\newblock \emph{\bibinfo{journal}{\mnras}} \textbf{\bibinfo{volume}{470}},
  \bibinfo{pages}{361--371} (\bibinfo{year}{2017}).
\end{thebibliography}

\noindent{\large\bf References}
\begin{thebibliography}{10}
\expandafter\ifx\csname url\endcsname\relax
  \def\url#1{\texttt{#1}}\fi
\expandafter\ifx\csname urlprefix\endcsname\relax\def\urlprefix{URL }\fi
\providecommand{\bibinfo}[2]{#2}
\providecommand{\eprint}[2][]{\url{#2}}

\makeatletter
\addtocounter{\@listctr}{47}
\makeatother

\bibitem{walton11}
\bibinfo{author}{{Walton}, D.~J.}, \bibinfo{author}{{Roberts}, T.~P.},
  \bibinfo{author}{{Mateos}, S.} \& \bibinfo{author}{{Heard}, V.}
\newblock \bibinfo{title}{{2XMM ultraluminous X-ray source candidates in nearby
  galaxies}}.
\newblock \emph{\bibinfo{journal}{\mnras}} \textbf{\bibinfo{volume}{416}},
  \bibinfo{pages}{1844--1861} (\bibinfo{year}{2011}).

\bibitem{predehl95}
\bibinfo{author}{{Predehl}, P.} \& \bibinfo{author}{{Schmitt}, J.~H.~M.~M.}
\newblock \bibinfo{title}{{X-raying the interstellar medium: ROSAT observations
  of dust scattering halos.}}
\newblock \emph{\bibinfo{journal}{\aap}} \textbf{\bibinfo{volume}{293}},
  \bibinfo{pages}{889--905} (\bibinfo{year}{1995}).

\bibitem{vaughan04}
\bibinfo{author}{{Vaughan}, S.} \emph{et~al.}
\newblock \bibinfo{title}{{The Discovery of an Evolving Dust-scattered X-Ray
  Halo around GRB 031203}}.
\newblock \emph{\bibinfo{journal}{\apjl}} \textbf{\bibinfo{volume}{603}},
  \bibinfo{pages}{L5--L8} (\bibinfo{year}{2004}).

\bibitem{vasilopoulos16}
\bibinfo{author}{{Vasilopoulos}, G.} \& \bibinfo{author}{{Petropoulou}, M.}
\newblock \bibinfo{title}{{The X-ray dust-scattered rings of the black hole
  low-mass binary V404 Cyg}}.
\newblock \emph{\bibinfo{journal}{\mnras}} \textbf{\bibinfo{volume}{455}},
  \bibinfo{pages}{4426--4441} (\bibinfo{year}{2016}).

\bibitem{mathis91}
\bibinfo{author}{{Mathis}, J.~S.} \& \bibinfo{author}{{Lee}, C.-W.}
\newblock \bibinfo{title}{{X-ray halos as diagnostics of interstellar grains}}.
\newblock \emph{\bibinfo{journal}{\apj}} \textbf{\bibinfo{volume}{376}},
  \bibinfo{pages}{490--499} (\bibinfo{year}{1991}).

\bibitem{draine03}
\bibinfo{author}{{Draine}, B.~T.}
\newblock \bibinfo{title}{{Scattering by Interstellar Dust Grains. II.
  X-Rays}}.
\newblock \emph{\bibinfo{journal}{\apj}} \textbf{\bibinfo{volume}{598}},
  \bibinfo{pages}{1026--1037} (\bibinfo{year}{2003}).

\bibitem{walton16}
\bibinfo{author}{{Walton}, D.~J.} \emph{et~al.}
\newblock \bibinfo{title}{{A 78 Day X-Ray Period Detected from NGC 5907 ULX1 by
  Swift}}.
\newblock \emph{\bibinfo{journal}{\apjl}} \textbf{\bibinfo{volume}{827}},
  \bibinfo{pages}{L13} (\bibinfo{year}{2016}).

\bibitem{israel17}
\bibinfo{author}{{Israel}, G.~L.} \emph{et~al.}
\newblock \bibinfo{title}{{An accreting pulsar with extreme properties drives
  an ultraluminous x-ray source in NGC 5907}}.
\newblock \emph{\bibinfo{journal}{Science}} \textbf{\bibinfo{volume}{355}},
  \bibinfo{pages}{817--819} (\bibinfo{year}{2017}).

\bibitem{xilouris99}
\bibinfo{author}{{Xilouris}, E.~M.}, \bibinfo{author}{{Byun}, Y.~I.},
  \bibinfo{author}{{Kylafis}, N.~D.}, \bibinfo{author}{{Paleologou}, E.~V.} \&
  \bibinfo{author}{{Papamastorakis}, J.}
\newblock \bibinfo{title}{{Are spiral galaxies optically thin or thick?}}
\newblock \emph{\bibinfo{journal}{\aap}} \textbf{\bibinfo{volume}{344}},
  \bibinfo{pages}{868--878} (\bibinfo{year}{1999}).

\bibitem{alton04}
\bibinfo{author}{{Alton}, P.~B.}, \bibinfo{author}{{Xilouris}, E.~M.},
  \bibinfo{author}{{Misiriotis}, A.}, \bibinfo{author}{{Dasyra}, K.~M.} \&
  \bibinfo{author}{{Dumke}, M.}
\newblock \bibinfo{title}{{The emissivity of dust grains in spiral galaxies}}.
\newblock \emph{\bibinfo{journal}{\aap}} \textbf{\bibinfo{volume}{425}},
  \bibinfo{pages}{109--120} (\bibinfo{year}{2004}).

\bibitem{watson11}
\bibinfo{author}{{Watson}, D.}
\newblock \bibinfo{title}{{The Galactic dust-to-metals ratio and metallicity
  using gamma-ray bursts}}.
\newblock \emph{\bibinfo{journal}{\aap}} \textbf{\bibinfo{volume}{533}},
  \bibinfo{pages}{A16} (\bibinfo{year}{2011}).

\bibitem{just06}
\bibinfo{author}{{Just}, A.}, \bibinfo{author}{{M{\"o}llenhoff}, C.} \&
  \bibinfo{author}{{Borch}, A.}
\newblock \bibinfo{title}{{An evolutionary disc model of the edge-on galaxy NGC
  5907}}.
\newblock \emph{\bibinfo{journal}{\aap}} \textbf{\bibinfo{volume}{459}},
  \bibinfo{pages}{703--716} (\bibinfo{year}{2006}).

\bibitem{rand96}
\bibinfo{author}{{Rand}, R.~J.}
\newblock \bibinfo{title}{{Diffuse Ionized Gas in Nine Edge-on Galaxies}}.
\newblock \emph{\bibinfo{journal}{\apj}} \textbf{\bibinfo{volume}{462}},
  \bibinfo{pages}{712} (\bibinfo{year}{1996}).

\bibitem{mezcua13}
\bibinfo{author}{{Mezcua}, M.}, \bibinfo{author}{{Roberts}, T.~P.},
  \bibinfo{author}{{Sutton}, A.~D.} \& \bibinfo{author}{{Lobanov}, A.~P.}
\newblock \bibinfo{title}{{Radio observations of extreme ULXs: revealing the
  most powerful ULX radio nebula ever or the jet of an intermediate-mass black
  hole?}}
\newblock \emph{\bibinfo{journal}{\mnras}} \textbf{\bibinfo{volume}{436}},
  \bibinfo{pages}{3128--3134} (\bibinfo{year}{2013}).
\newblock \eprint{1309.5721}.

\bibitem{weaver77}
\bibinfo{author}{{Weaver}, R.}, \bibinfo{author}{{McCray}, R.},
  \bibinfo{author}{{Castor}, J.}, \bibinfo{author}{{Shapiro}, P.} \&
  \bibinfo{author}{{Moore}, R.}
\newblock \bibinfo{title}{{Interstellar bubbles. II - Structure and
  evolution}}.
\newblock \emph{\bibinfo{journal}{\apj}} \textbf{\bibinfo{volume}{218}},
  \bibinfo{pages}{377--395} (\bibinfo{year}{1977}).

\bibitem{dopita96}
\bibinfo{author}{{Dopita}, M.~A.} \& \bibinfo{author}{{Sutherland}, R.~S.}
\newblock \bibinfo{title}{{Spectral Signatures of Fast Shocks. I. Low-Density
  Model Grid}}.
\newblock \emph{\bibinfo{journal}{\apjs}} \textbf{\bibinfo{volume}{102}},
  \bibinfo{pages}{161} (\bibinfo{year}{1996}).

\bibitem{allen08}
\bibinfo{author}{{Allen}, M.~G.}, \bibinfo{author}{{Groves}, B.~A.},
  \bibinfo{author}{{Dopita}, M.~A.}, \bibinfo{author}{{Sutherland}, R.~S.} \&
  \bibinfo{author}{{Kewley}, L.~J.}
\newblock \bibinfo{title}{{The MAPPINGS III Library of Fast Radiative Shock
  Models}}.
\newblock \emph{\bibinfo{journal}{\apjs}} \textbf{\bibinfo{volume}{178}},
  \bibinfo{pages}{20--55} (\bibinfo{year}{2008}).

\bibitem{heida19}
\bibinfo{author}{{Heida}, M.} \emph{et~al.}
\newblock \bibinfo{title}{{Searching for the Donor Stars of ULX Pulsars}}.
\newblock \emph{\bibinfo{journal}{\apj}} \textbf{\bibinfo{volume}{871}},
  \bibinfo{pages}{231} (\bibinfo{year}{2019}).

\bibitem{rybicki86}
\bibinfo{author}{{Rybicki}, G.~B.} \& \bibinfo{author}{{Lightman}, A.~P.}
\newblock \emph{\bibinfo{title}{{Radiative Processes in Astrophysics}}}
  (\bibinfo{publisher}{John Wiley \& sons, Inc.}, \bibinfo{year}{1986}).

\bibitem{caplan86}
\bibinfo{author}{{Caplan}, J.} \& \bibinfo{author}{{Deharveng}, L.}
\newblock \bibinfo{title}{{Extinction and reddening of H II regions in the
  Large Magellanic Cloud.}}
\newblock \emph{\bibinfo{journal}{\aap}} \textbf{\bibinfo{volume}{155}},
  \bibinfo{pages}{297--313} (\bibinfo{year}{1986}).

\bibitem{crane89}
\bibinfo{author}{{Crane}, P.~C.} \& \bibinfo{author}{{Napier}, P.~J.}
\newblock \bibinfo{title}{{Sensitivity}}.
\newblock In \bibinfo{editor}{{Perley}, R.~A.}, \bibinfo{editor}{{Schwab},
  F.~R.} \& \bibinfo{editor}{{Bridle}, A.~H.} (eds.)
  \emph{\bibinfo{booktitle}{Synthesis Imaging in Radio Astronomy}}, {\it
  Astron.~Soc.~Pac.~Conf.~Series} (\bibinfo{year}{1989}).

\end{thebibliography}
\end{document}